\documentclass[fleqn,10pt]{wlscirep}

\usepackage[utf8]{inputenc}
\usepackage[T1]{fontenc}

\usepackage{graphicx}
\usepackage{amssymb}
\usepackage{epstopdf}
\usepackage{url}
\usepackage{subfigure}
\usepackage{lineno}
\usepackage{color}

\usepackage{mathtools}
\usepackage{color, colortbl}
\usepackage{fancyhdr}
\usepackage{ctable}
\definecolor{Gray}{gray}{0.9}
\usepackage{enumitem}
\usepackage{graphicx}
\usepackage{caption}
\usepackage{color, colortbl}
\usepackage{array}
\usepackage{booktabs} 
\usepackage{makecell}
\usepackage{cellspace}
\usepackage{tabularx}
\usepackage{multirow}
\usepackage{amsmath}

\usepackage{xr}

\usepackage{verbatim}

\title{Pressure Test: Quantifying the impact of positive stress on companies from online employee reviews}

\author[1]{Sanja \v{S}\'{c}epanovi\'{c}}
\author[1]{Marios Constantinides}
\author[1,2,*]{Daniele Quercia}
\author[3]{Seunghyun Kim}
\affil[1]{Nokia Bell Labs, Cambridge, United Kingdom}
\affil[2]{CUSP, Kings College London, United Kingdom}
\affil[3]{Georgia Tech, USA}

\affil[*]{quercia@cantab.net}

\begin{abstract}
Workplace stress is often considered to be negative, yet lab studies on individuals suggest that not all stress is bad. There are two types of stress: \emph{distress} refers to harmful stimuli, while \emph{eustress} refers to healthy, euphoric stimuli that create a sense of fulfillment and achievement. Telling the two types of stress apart is challenging, let alone quantifying their impact across corporations. By leveraging a dataset of 440K reviews about S\&P 500 companies published during twelve successive years, we developed a deep learning framework to extract stress mentions from these reviews. We proposed a new methodology that places 
 each company on a stress-by-rating quadrant (based on its overall stress score and overall rating on the site), and accordingly scores the company to be, on average, either  a \emph{low stress}, \emph{passive}, \emph{negative stress}, or \emph{positive stress} company. We found that (former) employees of positive stress companies tended to describe high-growth and collaborative workplaces in their reviews, and that such companies' stock evaluations grew, on average, 5.1 times in 10 years (2009-2019) as opposed to the companies of the other three stress types that grew, on average, 3.7 times in the same time period. We also found that the four stress scores aggregated every year---from 2008 to 2020 ---closely followed the unemployment rate in the U.S.: a year of positive stress (2008) was rapidly followed by several years of negative stress (2009-2015), which peaked during the Great Recession (2009-2011). These results suggest that automated analyses of the language used by employees on corporate social-networking tools offer yet another way of tracking workplace stress, allowing quantification of its impact on corporations.
\end{abstract}
\begin{document}

\flushbottom
\maketitle
%
%
\thispagestyle{empty}


\section*{Introduction}

According to the American Institute of Stress, 40\% of workers consider their jobs to be stressful; a number that has significantly increased  during the COVID-19 pandemic~\cite{hbr_stress_cool}. The World Health Organization treats stress as the number one health threat in the U.S., with more than 60\% of doctor visits being due to a stress-related issue~\cite{nerurkar2013physicians}. Workplace stress is often linked to lower motivation, poor performance, and decline in employees' well-being~\cite{cartwright1997managing}, while it is estimated to amount to 190 billions in healthcare costs in the U.S. alone~\cite{hbr_meddl_stress}. To currently track how its employees deal with stress, a large company  would typically  recruit  consultants who would then administer surveys tailored to the company's situation, which typically end up being costly~\cite{vaske2011advantages},  
and restricted to a limited pool of self-selected participants~\cite{duda2010fallacy,gigliotti2011comparison,fricker2002advantages}. The current situation points to the need of more research. 

Stress is defined as ``a set of physical and psychological responses to external conditions or influences, known as stressors''~\cite{selye1956stress}. According to Lazarus~\cite{lazarus2000toward}, ``any change, either good (eustress) or bad (distress), is stressful, and whether it is a positive or a negative change, the physiological response is the same.'' To cope with workplace stress, an employee has to cognitively acknowledge that a situation causes stress before even attempting to manage it~\cite{colligan2006workplace}. Kobasa's framework of psychological hardiness offers three main categories of coping strategies~\cite{kobasa1979stressful}: \emph{commitment} (having an active involvement in one's own work with a sense of purpose), \emph{control} (believing and acting instead of feeling helpless in front of adversity), and \emph{challenge} (believing that change could be a source of improvement). Kobasa posited that these categories could help individuals face challenges and, as such, individuals could turn stressful events into opportunities for personal growth~\cite{kobasa1979stressful}. However, despite having explored the relation between stress and job performance for decades, researchers have not yet established whether stress and performance are in a negative linear relation, a positive linear relation, or  an inverted-U relation~\cite{lori03}. 

To tackle this gap, we draw upon two streams of previous literature, that is, literature about stress in the corporate context, and literature on how to gauge stress from online data. In the literature about stress in the corporate context, stress is often being portrayed as negative~\cite{wallis1983stress}; and as a leading cause of death, poor work performance, and diminishing well-being~\cite{cartwright1997managing}. More recently, however, researchers have advocated that there exist another type of stress: \emph{positive stress}. The idea is that whether stress is positive or negative depends on how an individual reacts to a stressor~\cite{crum2013rethinking}. \emph{`One's stress mindset can be conceptualized as the extent to which one holds the belief that stress has enhancing consequences for various stress-related outcomes such as performance and productivity, health and well-being, and learning and growth, or holds the belief that stress has debilitating consequences for those outcomes~\cite{crum2013rethinking}'}. Of prime importance is to distinguish appraisal from stress mindset. Stress mindset describes the evaluation of the nature of stress itself as positive or negative (i.e., enhancing or debilitating)~\cite{crum2013rethinking}, whereas appraisal is about the evaluation of a particular stressor as more or less stressful~\cite{cohen1983global}. For example, one may appraise a difficult task as highly stressful and have a stress debilitating mindset, which, in turn, leads the individual to experience the situation as draining (negative stress). By contrast, another individual may again consider the task as highly stressful but have a stress enhancing mindset, leading the individual to experience the situation as an opportunity for growth and development (positive stress). While stress is often linked to depression~\cite{hammen2005stress, wang2005work}, several accounts posit that certain stressful experiences may fundamentally change individuals for the better---a phenomenon referred to as stress-related growth~\cite{crum2013rethinking}. The experience of stress can enhance the development of mental toughness, greater appreciation for life, and an increased sense of meaningfulness~\cite{park2006introduction, tedeschi2004posttraumatic}. However, as Crum et al.~\cite{crum2013rethinking} pointed out, these conflicting findings in the stress literature suggest a nuanced view of stress. A view that recognizes the debilitating nature of stress on health and performance, but can also account for its enhancing nature in specific circumstances. We hypothesized that the presence of both positive and negative stress can be measured from digital data based on previous literature that has done just that with different techniques upon different datasets. Guntuku et al.~\cite{guntuku2019understanding} used the Linguistic Inquiry and Word Count (LIWC)~\cite{pennebaker2001linguistic} dictionary's features (e.g., topical categories, emotions, parts-of-speech) to predict stress of social media (Facebook and Twitter) users. Saha and De Choudhury~\cite{saha2017modeling} did a similar study but on Reddit and did so in conjunction with gun violence events, and found specific stress markers to be associated with linguistic changes about \emph{``higher self pre-occupation and death-related discussion.''} Similar to our study, Vedant et al.~\cite{das2020modeling} showed that the use of language in employee reviews can be used to operationalize organizational culture: the collection of values, expectations, and practices that guide and inform employees' actions within a company.

Based on these preliminary findings, we hypothesized that workplace stress is reflected in company reviews. To explore this hypothesis, we placed companies on a 2x2 coordinate system, based on their overall stress scores and overall ratings on the company review site. This stress-by-rating quadrant  effectively divided companies in four stress types that we termed low stress, passive, negative stress, and positive stress (Table \ref{tab:example_reviews} shows example reviews of companies of each stress type). Low stress companies enjoy high overall ratings and low stress scores. These are usually established organizations that offer workplace flexibility, good pay, and bonuses. Passive companies are characterized by low overall ratings and a small proportion of posts mentioning stress. They tend to have high turnover, due to repetitive workload and non-motivated employees.  Negative stress companies are characterized by low overall ratings and a high proportion of posts mentioning stress. Employees of these companies are particularly unhappy as, in addition to the unsatisfactory conditions, they also experience high pressure. Finally, positive stress companies enjoy high overall ratings but also high stress scores. These tend to be inspiring, reputable workplaces that attract employees because of the collaborative atmosphere and career prospects despite the pressure employees are subject to. The project website for our study is found on \url{https://social-dynamics.net/positive-stress}.

\subsection*{Data}
\label{sec:section4}
After obtaining the U.S. unemployment rates between 2008 and 2020 from the U.S. Bureau of Labor Statistics~\cite{bls} and the S\&P 500 stock market data (including the 500 large capital U.S. companies with a cumulative market capitalization to be around 70-80\% of the total in the country) from the Yahoo Finance portal~\cite{yahoo}, we matched that data with our company reviews. More specifically, we obtained 440K geo-referenced posts on Glassdoor (\url{https://www.glassdoor.com}), a popular company reviewing site about the S\&P 500 companies published during twelve successive years, from 2008 to 2020. On this site,  current and, more likely, former employees of companies write reviews about their own corporate experience, ranging from job interviews to salaries to workplace culture. The site provided an overall rating of each company based on  employees' reviews. As of 2021, there are 50M monthly visitors on the platform, and 70M reviews about 1.3M companies. To ensure quality reviews, the site employs the following three mechanisms. First, an automatic (proprietary) and manual content moderation system is paired with the possibility for users to flag content. Such a combined system partly prevents fake reviews (e.g., a company unfairly forcing employees to leave positive reviews). Second, every user wanting to browse others' content has to take the time to write one review. This requirement encourages the presence of neutral reviews and  partly prevents the so-called \emph{non-response bias}, a situation in which the opinions of respondents are very different from those of non-respondents. Third, the site allows for a maximum of one review per employee per company per year, preventing any employee from contributing a disproportionate number of reviews, and, in so doing, discouraging \emph{sampling bias}, a situation in which the opinions of some members are more represented in the data than those of others.

\begin{table}[ht!]
\footnotesize
\caption{Example reviews of companies of each stress type.}
\label{tab:example_reviews}
\centering
\begin{tabular}{p{0.18\textwidth}p{0.7\textwidth}}
\toprule
        \textbf{Stress Type} & \textbf{Review excerpt}  \\
\midrule
\rowcolor{Gray} Low stress & My company walks its talk. It \emph{[the company]} takes care of customers and employees.\\
Negative stress & There is a feeling of scarcity due to the constant reorganizations, pressure, and surprise layoffs. \emph{[...]} You could imagine how toxic the environment is.     \\
\rowcolor{Gray} Passive stress  &  There is no regard for how the remaining work will get done, just how the bottom line looks at that moment in time. People are not treated as respected contributors to the organization.  \emph{[...]} This is a very unstable, unhealthy, volatile, stressed out environment, with incredibly poor leadership.   \\
Positive stress & You have to be a very driven and self-motivated person to be successful here. If you are willing to commit and put in the extra effort and hard work, it will be extremely worth it. \emph{[...]}  Every day is very busy and it can be stressful at times but its all worth it!.\\
\bottomrule
\end{tabular}
\end{table}
\section*{Methods}

\label{sec:section5}

\begin{figure*}
   \centering
    \includegraphics[width=.97\linewidth]{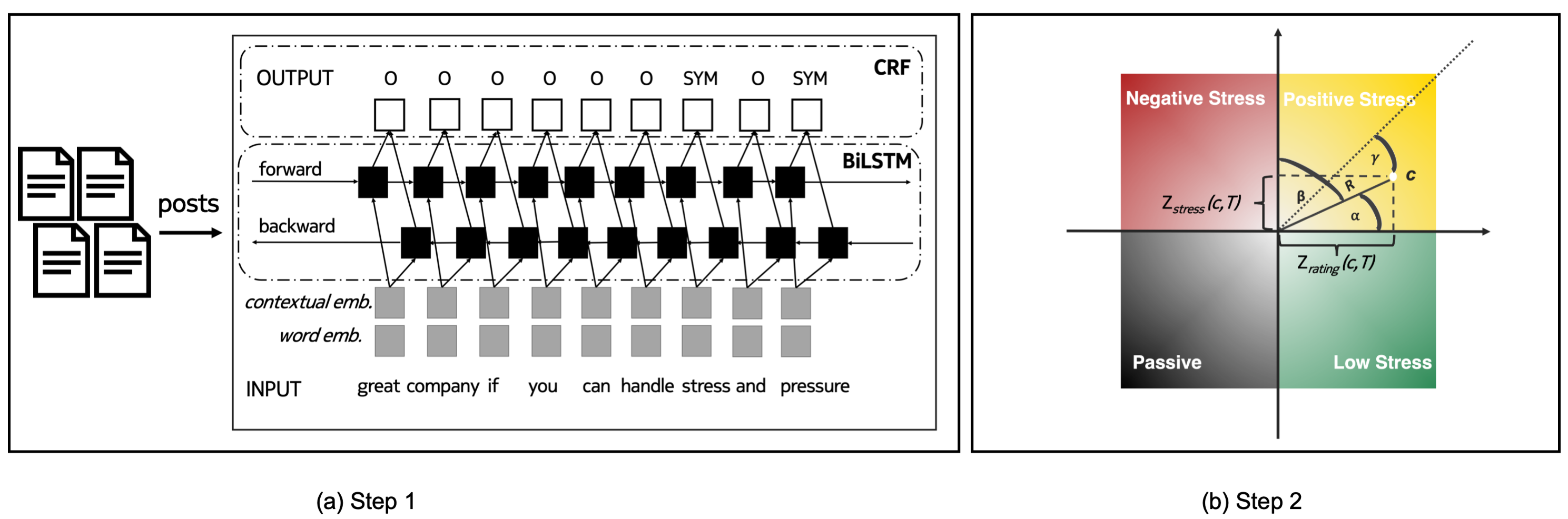}
    \caption{Placing companies on a stress-by-rating quadrant by  detecting stress mentions in reviews about a company using a state-of-the-art NLP deep-learning framework (\emph{Step 1}),  placing the company in the rating-by-stress quadrant, and computing its association with its stress type (i.e., with its quadrant)  (\emph{Step 2}). To see how the association is computed, consider company $c$ shown in \emph{(b)} to be of positive stress. $c$ is placed according to its $z_{rating}(c,T)$ along the $x$-axis, and to its  $z_{stress}(c,T)$ along the  $y$-axis. $R$ is the radius from the center to $c$'s point; $\alpha$ is the angle between the radius line and the $x$-axis; $\beta$  is the angle between the radius line and the  $y$-axis; and $\gamma$ is the angle between the radius line and the diagonal shown as a dotted line. The function $f(c,s,T)$ combines $R$, $\alpha$, $\beta$, and $\gamma$, and accordingly scores  $c$ to have a  high association weight with positive stress $s$ during period $T$ (darker shades of colors), as $c$ is close to the quadrant's diagonal, and distant from the intersection point.  }
    \label{fig:quadrant}
\end{figure*}

We extracted  mentions related to stress using an NLP deep-learning tool, which  is trained to extract medical conditions from  free-form text~\cite{} (Figure~\ref{fig:quadrant}(a)). We then designed a new methodology that placed the 500 S\&P companies on a stress-by-rating quadrant  based on their overall ratings on the reviewing site on one axis, and the presence of mentions related to stress in their reviews on the other axis (Figure~\ref{fig:quadrant}(b)). In so doing, we  classified  each company to be, on average, either  a \emph{low stress}, \emph{passive}, \emph{negative stress}, or \emph{positive stress} company. We finally computed each company's strength of membership to its quadrant depending on whether the company is placed both close to the diagonal and far from the (0,0) intersection point. The function $f(c,s,T)$, which is expressed in Equation~(\ref{eq:q_w}) and graphically depicted in Figure~\ref{fig:quadrant}(b), assigns a higher weight to company $c$ of stress type $s$, if $c$ is both closer to the quadrant's diagonal (i.e., it is farther from the remaining quadrants) and more distant from the two axes' intersection (i.e., it has higher absolute overall rating and stress score values). We  call $f(c,s, T)$ to be company $c$'s association with stress type $s$ during $T$ since the higher its value, the more $c$ is associated with stress type $s$ during $T$.

\mbox{ } \\
\noindent
\textbf{Extracting stress mentions from posts.} 
To extract stress mentions, we used the MedDL entity extraction module~\cite{scepanovic2020extracting} (the left rectangle in Figure~1 (a)). MedDL uses \emph{contextual embeddings} and a \emph{deep BiLSTM-CRF sequence labeling architecture} (we used the default parameters as specified in~\cite{scepanovic2020extracting}).
The model was pre-trained  and evaluated on a labeled dataset of Reddit posts called MedRed~\cite{scepanovic2020extracting}. The MedRed dataset was split into train (50\%), dev (25\%), and test (25\%) sets. We evalued MedDL using the strict and relaxed $F1$-scores, two commonly used performance metrics for entity extraction models. The \emph{strict $F1$-score} counts only the \emph{exact} matches with the true labels as correct, while the \emph{relaxed $F1$-score} takes into accoun the partially matching extracted entities. We provide formulae for the two scores in \emph{Supplementary Information} (SI Equation (1)). MedDL was compared against two well-known entity extraction tools: MetaMap (a well-established tool \cite{aronson2010overview} and a de-facto baseline method for NLP studies related to health \cite{tutubalina2018medical}) and TaggerOne (a machine learning tool using semi-Markov models and a medical lexicon \cite{leaman2016taggerone}). The MedDL method achieved a strict/relaxed $F1$-score of $.71$/$.85$ when extracting symptoms (Figure S4), outperforming both MetaMap and TaggerOne by a large margin (the two have $F1$-scores of $.17/.48$ and $.31/.58$, respectively). Furthermore, MedDL has shown generalizability when applied on dataset  (e.g., dream reports \cite{vscepanovic2022epidemic}) different than those it was trained on (i.e., Reddit data).

\begin{table}[t] 
\small
\caption{Top15 most frequent stress-related mentions identified on a company review site, and their frequency counts.}
\begin{tabular}{lll}
\hline
{Condition related to stress} & {\# mentions} & {Example mention}\\ \hline
stress & 3473 & \emph{``Great company to work for, if you can handle stress.''} \\ 
high stress & 710 & \emph{``High stress work environment, long work hours.''} \\ 
pressure & 447 & \emph{``a lot of pressure to get things done.''} \\ 
burnout & 277  & \emph{``[...], the ones who made the cut to stay are suffering from burnout.''}\\ 
understaffing & 99 & \emph{``Somewhat job stability due to understaffing.''} \\ 
heavy workload & 58 & \emph{``Lack of work/life balance, extremely heavy workload.''} \\ 
exhaustion & 58  & \emph{``You will be pushed to the point of exhaustion [...].''}\\
stress levels & 57 & \emph{``[...] stress levels peak insanely when the store manager [...].''} \\
overworked & 45 & \emph{``At times, you can feel overworked and undervalued.''} \\
tension & 38 & \emph{``There's a lot of tension between coworkers because of commission.''} \\
high workload & 38 & \emph{``[...] seeing many large set-backs which cause very high workload''} \\
extreme stress & 33 & \emph{``Beware: extreme stress and pressure.''} \\
mental stress & 23 & \emph{``[...] ends up giving you a lot of mental stress.''}\\
overload & 17 & \emph{``No work life balance [...], overloaded and benefits are not good.''} \\
pressure to perform & 9 & \emph{``[...] a lot of pressure to perform, long working hours''} \\
\hline
\end{tabular}
\label{table:stress-related_symptoms}
\end{table}

We extracted stress mentions in three steps (further detailed in \emph{Supplementary Information}). First, we detected over 21K posts that mentioned over 5K unique medical conditions. Most frequent medical conditions identified include \emph{stress}, \emph{pain}, \emph{headache}, and \emph{depression}. Second, we inspected the $\textrm{top 200}$ most mentioned conditions and manually selected $31$ of them that specifically reflect  workplace stress ($\textrm{top 15}$ are shown in Table~\ref{table:stress-related_symptoms}). Third, we extracted all reviews mentioning any of the $31$ conditions. This resulted in $7,338$ posts related to stress, which accounted for $1\%$ of all posts. Despite this seemingly low number of posts, when aggregated, these posts returned statistically significant results for our metrics, which are described next.

\mbox{ } \\
\noindent
\textbf{Associating stress types with companies.} We placed each S\&P 500 company on a stress-by-rating quadrant. More specifically, for each company $c$, we computed its average review rating and its stress score: 
\begin{align}
\textit{rating(c,T)} = \textit{$c$'s average review rating during } T, \nonumber \\
\textit{stress(c,T)}=
 \frac{ \textit{\# $c$'s posts related to stress during } T } { \textit{total \# $c$'s posts during } T}. \nonumber
\end{align}
where  $T$ is set to initially include all the years under study (2009-2019). To then ease comparability, we $z$-scored these two values:
\begin{align}
z_{rating}(c,T) & =  \frac{rating(c,T) - \mu_{rating}(T)}{\sigma_{rating}(T)},  \nonumber \\
z_{stress}(c,T) & =  \frac{stress(c,T) - \mu_{stress}(T)}{\sigma_{stress}(T)}. \nonumber
\end{align}
where $\mu_{rating}(T)$ and $\sigma_{rating}(T)$ are the average and standard deviation of the review ratings for all companies (regardless of their stress types) during the whole period $T$ (readily available on the company review site), and $\mu_{stress}(T)$ and $\sigma_{stress}(T)$  are the average and standard deviation of the stress scores for all companies during the whole period $T$.

\vspace{10pt}
Each S\&P 500 company was assigned to one of the four quadrants based on the signs of their two z-scores (Figure~\ref{fig:quadrant}(b)). For example, a company $c$ with a negative  $z_{rating}(c,T)$ and a positive $ z_{stress} (c,T)$ would be placed in the negative stress quadrant, while a company with a positive $z_{rating}(c,T)$ and a positive $ z_{stress} (c,T)$ would be placed in the positive stress quadrant. The resulting quadrants are consequently four:

\begin{itemize}[leftmargin=*, label={}]
    \item \emph{Low Stress  {companies}} {- These enjoy} high overall ratings and low stress scores. Their employees tended to think very positively about their workplace experience with comparatively fewer posts {mentioning stress conditions}.
    \item \emph{Passive  {companies}} -- 
    {These are characterized by} low overall ratings and {a small proportion of posts mentioning stress}. Their employees were mostly not satisfied with their jobs, but they also showed comparatively fewer signs of stress in their reviews. 
    \item \emph{Negative stress {companies}} --
    {These are characterized by} high stress scores and low overall ratings. Their employees mentioned stress conditions, while also scoring their workplace experience low.
    \item \emph{Positive stress {companies}} -- 
    {These enjoy}  high ratings {despite} high stress scores. Their employees mentioned stress yet did so in the context of  high-pressure and highly rewarding work environments.
\end{itemize}

Once a company $c$ is placed in its quadrant (i.e., associated with its stress type $s$), we needed to estimate its association with this quadrant, i.e., with $s$. For example,  company $c$ with ($z_{rating} (c,T), z_{stress}(c,T)$) equal to (3,3) is more strongly associated with \textit{positive stress}, than what a company with $(0.5,0.5)$ would be. To estimate $c$'s association with $s$, we combined $c$'s two $z$-scores concerning review rating and stress score as follows (and as depicted in Figure~\ref{fig:quadrant}(b)):

\begin{align}\label{eq:q_w}
f(c,s, T)&=\left\{
  \begin{array}{@{}ll@{}}
    l  ( z_{rating}(c,T), z_{stress}(c,T) ) = R / (\gamma + \pi), & \text{if}\  c \in s \text{ during } T;\\
    0, & \text{if}\  c \notin s \text{, or } $c$ \text{ has no review during } T;
  \end{array}\right.\\
 \text{where:} \nonumber \\
    R &= \sqrt{ z_{rating}(c,T) {}^2 + z_{stress}{}(c,T)^2}, \nonumber  \\ 
    \gamma &= max ( (\alpha - \pi / 4), (\beta - \pi / 4)),  \quad \quad 
    \nonumber \\
    \alpha &= arccos(|z_{rating}(c,T){})| / R),  \quad \quad \quad \quad \quad 
    \nonumber \\
     \beta &= arccos(|z_{stress}(c,T){})| / R). \nonumber
\end{align}
where $T$ is initially set to include all the years under study, from 2009 to 2019. To ease understanding of the above formula, consider that function $l$, on input of the two $z$-scores (i.e., the company's two coordinates in the quadrant), computes the extent to which  company $c$ is on the diagonal and far from the (0,0) intersection point (Figure~\ref{fig:quadrant}(b)). It gives higher weights to companies that are both closer to the quadrant's diagonal (i.e., which are farthest from the remaining quadrants) and more distant from the  axes' intersection  (i.e., which have higher absolute rating/stress score values).

\mbox{ } \\
\noindent
\textbf{Computing stress scores over the years.} For each year $y$, we quantified the amount of a given stress type $s$ expressed in the posts produced in that year. More specifically, we computed:
\begin{align}\label{eq:q_temp_aggr}
    	m{(s,y)} = 
    	\sum_{c \in s} f(c,s,y) \times w{(c,y,s)},
\end{align}
For all the companies of stress type $s$, we summed each company's association $f(c,s,y)$ with $s$ during year $y$ weighted by 
 the presence of posts about the company during  $y$ (giving higher weights to companies whose employees contributed more reviews in that year):
\begin{equation}
  w(c,y,s)=\left\{
  \begin{array}{@{}ll@{}}
    \frac{ \textit{\# $c$'s posts in year $y$} } { \textit{total \# posts in year $y$}}, & \text{if}\  c \in s \textit{ in year } y;\\
    0, & \text{if $c$ has no reviews in year $y$.}
  \end{array}\right.
\end{equation}

\mbox{ } \\
\noindent
\textbf{Associating topical categories with stress types.} 
To identify relevant words for each stress type, we run BERTopic~\cite{grootendorst10bertopic}, which is a state-of-the-art topic modeling algorithm. A topic modeling algorithm is an unsupervised technique to extract topics that appear frequently in a piece of text (in our case,  a post). The algorithm works in four sequential steps:

\begin{enumerate}
    \item converts each post into a 512-dimensional vector (called embedding) of numerical values using a pre-trained BERT-based sentence transformer~\cite{devlin2018bert} (in our case, we used the default model, that is, the ``paraphrase-MiniLM-L6-v2''). BERT (Bidirectional Encoder Representations from Transformers) is a state-of-the-art transformer-based machine learning technique for natural language processing (NLP), which takes into account the context of each word. 
    \item reduces dimensionality using UMAP~\cite{mcinnes2018umap} (or Unification Map) for every embedding, as many clustering algorithms handle high dimensionality poorly. UMAP is arguably the best performing dimensionality reduction algorithm as it keeps significant portion of the high-dimensional structure in lower dimensionality. 
    \item uses HDBSCAN~\cite{mcinnes2017hdbscan} for clustering with the ``UMAP'' embeddings, resulting in similar posts being clustered together. HDBSCAN is a density-based algorithm that works well with UMAP as the structure is preserved in a lower-dimensional space. Additionally, HDBSCAN does not force data points to clusters as it considers them outliers.
    \item identifies keywords using the c-TF-IDF~\cite{grootendorst10bertopic} score (Equation~\ref{eq:ctfidf}), and using that score, derives topics from the identified keywords. To create a topic representation, we took the top 3 keywords per topic based on their c-TF-IDF scores. The higher the score, the more representative is as the score is a proxy of information density.
    \begin{equation}
    \label{eq:ctfidf}
        \textrm{c-TF-IDF}_{l}  = \frac{k_{l}}{o_{l}} \times \frac{p}{\sum_{j}^{q}k_{j}}
    \end{equation}
where the frequency of each keyword $k$ is extracted for each topic $l$ and divided by the total number of keywords $o$. The total, unjoined, number of posts $p$ is divided by the total frequency of keyword $k$ across all topics $q$.
\end{enumerate}

\subsection*{Analysis plan}
Our analysis plan unfolded in three steps. First, as an initial validation step, we  ascertained that stress was paraphrased in a company's reviews differently according to the company's stress type. Second,  we tested whether the evolution of each stress score over the years tallied with large-scale exogenous events such as the Great Recession. Third, we tested that a company's stress type is partly associated with its  stock growth.
\section*{Results}
\subsection*{Topics discussed in reviews of companies of different stress types}
To ascertain whether the content of the reviews captured aspects specific to the four stress types, we identified the top relevant words for each type by running a topic modeling algorithm called BERTopic~\cite{grootendorst10bertopic}, and did so  on four distinct sets of reviews: each set contained reviews of all the companies of a given stress type. This algorithm found the emergent topics in the four sets, and Table~\ref{table:linguistic_Q_new} lists the  the top three words for each topic. The top10 topics for each quadrant are statistically associated with the quadrant. That is, based on chi-square tests, each topic $l$ associated with quadrant $s$: has frequency  in $s$ always above zero (is dependent on $s$), and is independent of any quadrant other than $s$. As detailed in \emph{Supplementary Information},
by inspecting these groups of words and corresponding representative reviews,  six annotators identified the emergence of three workplace themes:

\begin{description}
\item  \emph{Career drivers} (first set of rows in Table~\ref{table:linguistic_Q_new}).  Negative stress companies were associated with words such as `overtime', `mandatory, `shifts', and the typical workplace described in the reviews, according to our annotators, was characterized by considerable emotional pressure. On the other hand, passive companies were associated with words such  as `vacation', `pto', and `vacation/sick', and the corresponding reviews tended to  deflect from the day-to-day  work and  focus on activities outside work such as vacation and time off.  Low stress companies were associated with words such as `scheduling', `flexibility', and `autonomy', and the typical workplace described in the reviews was one in which employees cherished their sense of control over their work. Finally, positive stress companies were associated with words such as `teamwork', `supportive', and `collaborative', and the typical workplace in the reviews was one with a collaborative and supportive  culture. 
 
\item  \emph{Industry or benefits} (second set of rows in Table~\ref{table:linguistic_Q_new}). Negative stress companies were  associated with words such as `discounts', `sale', `coupons',  while positive stress companies were associated with words such as `gain', `billions', and `software'. Their reviews were effectively mentioning the  industry sectors they referred to: Consumer Discretionary (e.g., retail shops) for the reviews of negative stress companies, and Information Technology for those of positive stress ones. On the other hand, passive companies were associated with words such as `insurance',  `espp', and `hsa', and, similarly, low stress ones  with words such as `401k', `bonus', and `retirement'; the corresponding reviews indicated workplaces in which concerns about long-term financial benefits rather than the presence of implicit incentives in one's own work were at the forefront.  

\item \emph{Emotional Aspects} (third set of rows  in Table~\ref{table:linguistic_Q_new}). Negative stress companies were associated with words such as `horrible', `terrible', and `awful', confirming, once again, the presence of  emotional pressure. Passive companies were instead associated with words such as `repetitive', `turnover', and `workload', confirming the tedious nature of those workplaces.  Low stress companies were associated with words such as `fair', `friendlygood', and `pays', and the corresponding reviews described a  good work-life balance. Finally, positive stress companies were associated with words such as `prestige', `boost', and `reputation', and their reviews 
described high performing, dynamic, and fast-paced workplaces.
\end{description}

{\def\arraystretch{1.5}
\begin{table*}[t!]
\setlength{\tabcolsep}{1.5mm}
\centering
\footnotesize
\begin{tabular}{c p{37mm} p{37mm} p{37mm}  p{37mm}}
\specialrule{.1em}{.05em}{.05em} 

 & \multicolumn{1}{l}{\textbf{Negative stress}} & \multicolumn{1}{l}{\textbf{Passive}} & \multicolumn{1}{l}{\textbf{Low stress}} &
 \multicolumn{1}{l}{\textbf{Positive stress}}\\
\Xhline{2\arrayrulewidth}
\multirow{3}{*}{\rotatebox[origin=c]{90}{\textbf{{\shortstack[c]{Career\\drivers}}}}} 

& overtime & vacation & scheduling   & teamwork  \\
& mandatory  & pto  & flexibility   & supportive \\
& shifts & vacation/sick & autonomy    & collaborative  \\

\specialrule{.1em}{.05em}{.05em} 
\multirow{3}{*}{\rotatebox[origin=c]{90}{\textbf{{\shortstack[c]{Industry\\or benefits}}}}} 

& discounts  & insurance & 401k & gain  \\
& sale & espp  & bonus  & billions  \\
& coupons  & hsa & retirement  & software  \\

\specialrule{.1em}{.05em}{.05em} 
\multirow{3}{*}{\rotatebox[origin=c]{90}{\textbf{{\shortstack[c]{Emotional\\ aspects}}}}} 

& horrible & repetitive & fair & prestige \\
& terrible  & turnover & friendly/good   & boost  \\
& awful   & workload  & pays & reputation  \\

\specialrule{.1em}{.05em}{.05em}

\end{tabular}
\caption{Three-word groups present in the reviews of companies of the four stress types. These groups were automatically found by BERTopic and speak to three main workplace characteristics: career drivers, industry and benefits, and emotional aspects. For each group, the top three words are shown together with their normalized word importance. Abbreviations of words describing monetary benefits include pto  (paid time off); espp (employee stock purchase plan); hsa (health savings account); 401k (a retirement savings and investing plan that employers offer).}
\label{table:linguistic_Q_new}
\end{table*}
}

\subsection*{Evolution of stress types and the Great Recession}
\begin{figure*}[t!]
    \centering
        \includegraphics[width=.82\linewidth]{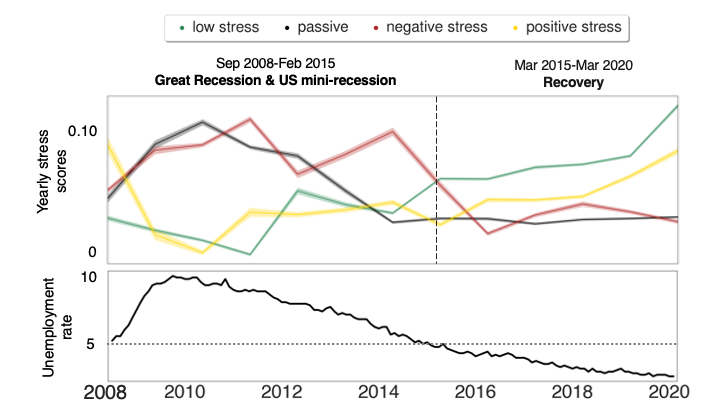} 
    \caption{The evolution of: \emph{(top)} the four types of stress; and \emph{(bottom)} the unemployment rate in the U.S., with  the horizontal dashed line reflecting pre-recession rate. The stress score per year  is calculated using Equation~(\ref{eq:q_temp_aggr}), and its standard deviations are shown with shaded lines.}
    \label{fig:stress_time}
 \end{figure*}

After the preliminary validation step in which we  ascertained that stress was paraphrased in reviews  differently according to the stress type, we tested whether the evolution of each stress score over the years tallied with large-scale exogenous events such as the Great Recession. We plotted the amount $m(s,y)$  of each stress score $s$ in each year $y$ (as per Equation~(\ref{eq:q_temp_aggr})), from 2008 to 2020 (top panel in Figure~\ref{fig:stress_time}).  The overall changes closely followed the unemployment rates from the U.S. Bureau of Labor (bottom panel in Figure \ref{fig:stress_time}): a year of positive stress (2008) was rapidly followed by several years of negative stress (2009-2015), which peaked during the Great Recession (2009-2011) during which  the U.S. went through a loss of over 7.5 million jobs and high unemployment rates~\cite{grusky2011consequences}.

\begin{figure}
	\begin{center}
		\centering
		\begin{minipage}[b]{0.32\linewidth}
		    \centering
		    \includegraphics[width=1.0\linewidth]{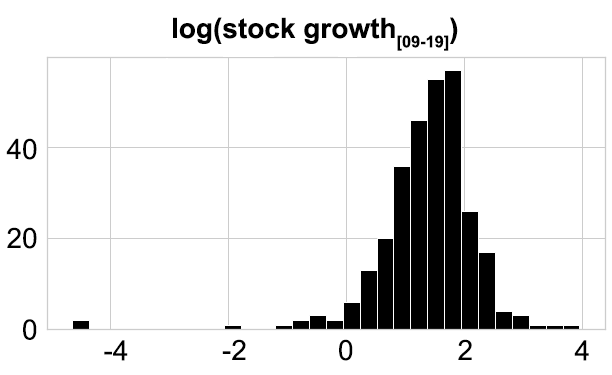}
		    \centerline{\small(a)}
		\end{minipage}
		\begin{minipage}[b]{0.64\linewidth}
		    \centering
		    \includegraphics[width=1.0\linewidth]{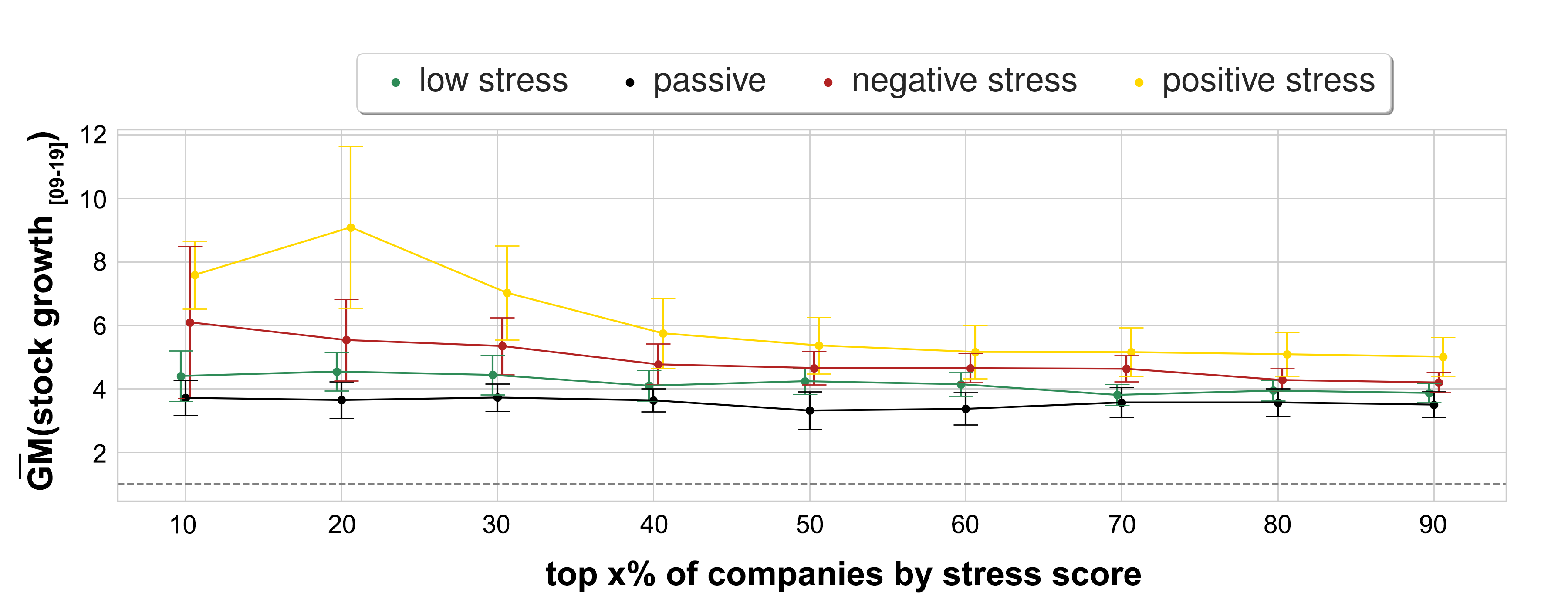}
		    \centerline{\small(b)}
		\end{minipage}
	\end{center}    
	\caption{\emph{(a)} Distribution across companies of the logarithm of stock growth values from the average stock price in 2009 and that of 2019 (${stock\_growth}_{[09-19]} = stock_{2019}/stock_{2009}$) showing the stock growth is log-normally distributed. The average stock price for year $y$ ($stock_y$) is calculated as the average of the daily Adjusted Closing Prices for the year. \emph{(b)} Geometric mean of the stock growth values $\bar{GM}({stock\_growth}_{[09-19]})$ for increasing stress score percentiles for the companies of a given stress type. Error bars represent geometric standard error $GSE({stock\_growth}_{[09-19]}) =$ $\bar{GM}({stock\_growth}_{[09-19]})/$ $\sqrt{N} \cdot \sigma(log({stock\_growth}_{[09-19]}))$.}
	\label{fig:stock_diff}
    \vspace{8pt}
\end{figure}

\subsection*{Stock growth of companies of different stress types}
Finally, we hypothesized that a company's way of dealing with workplace stress was partly \emph{associated} with  performance. Given our data, we cannot study whether  stress \emph{causes} (poor) performance but can only study how the two are associated. Also, there is no company performance measure that is solely affected by a company's stress culture. As such, our stress scores are unlikely to be predictive of any company-wide performance measure.  We opted for long-term stock growth as our performance measure, not least because it is publicly available and standardized across companies. However, such a growth is partly affected by a company's culture, and conflates endogenous factors (e.g., productivity) with exogenous ones (e.g., financial cycles). Yet we expected  our stress measures to qualitatively describe different forms of financial success, at least in the long term. To that end, we computed stock growth during the full period of study, that is, between 2009 to 2019:
\begin{equation}
\label{eq:stock_main}
\textrm{stock growth}_{[09-19]}(c)=  \frac{stock(c)_{2019}}{stock(c)_{2009}}
\end{equation}
where $stock^{i}$ is the average adjusted closing price of company $c$'s stock in year $i$. We chose long-term growth instead of short-term one (e.g., that pertaining 2018-2019) to partly account for any potential influence of exogenous events (e.g., Great Recession, market manipulation, incidental growth/decline~\cite{forbers21}). In \emph{Supplementary Information}, we show that the results do not qualitatively change  when considering the narrower 5-year period from 2014 to 2019.
Given a stress type $s$, we computed company $c$'s  \emph{association} $f(c,s, T)$ with $s$ during time period $T$ (initially set to the whole period of study), consequently grouping all the companies of a given stress type into their stress score percentiles (Figure \ref{fig:stock_diff}b).  As the distribution of stock growth values across companies is heavy-tailed (Figure \ref{fig:stock_diff}a), we used the geometric mean to average these values across companies. That is, $GM({\textrm{stock growth}_{[09-19]}}) = \Pi (\textrm{stock growth}_{[09-19]}(c))^{1/n} $, where $c$ is a company in a specific \emph{(stress type,percentile)} bin, and $n$ is the number of the companies in such a bin. Positive stress companies enjoyed the highest stock growth with an average value across all percentiles of $\bar{GM}(\textrm{stock growth}_{[09-19]}) = 5.07$ (Figure \ref{fig:stock_diff}b), while the average stock growth across the other three types of companies was noticeably lower ($\bar{GM}({\textrm{stock growth}_{09-19}}) = 3.70$), with passive stress companies exhibiting the lowest growth ($\bar{GM}({\textrm{stock growth}_{09-19}}) = 3.42$). To ease the interpretation of such values, consider the example of Equinix, a digital infrastructure company headquartered in California, which our approach labeled to be a ``positive stress'' company. Its  stock price traded at 61\$ in 2009 and its stock price climbed over 695\% (i.e., its $\bar{GM}(\textrm{stock growth}_{[09-19]})$ was 7.95), trading at 485\$ ten years later. 
\section*{Discussion}

\subsection*{Limitations}
This work has five main limitations. The first concerns the inability of studying whether stress causes performance differences given the absence of cross-lag data that links performance to a stress-related company-wide indicator. Theoretically, we could run a lagged analysis as a linear regression where the dependent variable is the company's growth at different time intervals. However, such an analysis is hard because of two main reasons: \emph{a)} no fine-grained temporal granularity for reviews is possible as reviews might be temporally misaligned since they could be posted after an employee leaves the company, and \emph{b)} many, mostly smaller, companies have joined the public reviewing site at later points in time, thus reviews will not cover all 12 years of analysis.

The second limitation is that the decreasing trend of stock growth may be dependent on the two main aspects: company ratings and industry sector. These two have little to do with the hypothesized relationship between stress and performance. We therefore repeated our analyses by considering a company's overall website rating and its industry sector. As for ratings, we indeed found increasing stock growth with increasing review ratings; still, positive stress companies experienced the highest growth (Figure S6 in \emph{Supplementary Information}) compared to highly-rated companies. As for industry sectors, we showed that tech companies were over-represented in the positive stress set, and stock growth was partly driven by them (Figure S7 in \emph{Supplementary Information}). However, by separating companies by industry sector, we still observed that positive stress companies grew more than the other three types (Figure S8 in \emph{Supplementary Information}). 

The third limitation concerns data biases related to temporal granularity and geographic representativeness. Upon new available data, future studies could study workplace stress outside US, allowing for cross-cultural comparisons.

The fourth limitation has to do with nuances when rating a company (e.g., being satisfied with the use of the overall company rating and not its composing dimensions). While on Glassdoor there are several rating fields available, only the overall rating field was mandatory and hence provided sufficient coverage for our analysis.

The fifth limitation is that the deep-learning model used to detect stress mentions in posts is not always accurate. Our medical entity extraction model has two main limitations. First, the model's strict/relaxed accuracy is .71/.85, and, even though it outperformed competitive baselines by a large margin, it still is not perfect. To partly address this issue, our method limits itself to textual mentions pertaining stress at work. Second, entity extraction models such as ours are not always able to tell apart personal from figurative health mentions (e.g., \emph{`I felt pain' vs. `He was such a pain to work with'}). This is still an active area of research. Yet our model is relying on a large transformer model (e.g., contextual embeddings RoBERTa), and, as such, it is less likely to make such errors than a simpler, keyword-matching technique. Future studies could use some of the newly published social media datasets~\cite{naseem2022identification} to further train our model to distinguish between different \emph{types of health mentions}.

\subsection*{Implications}

To place our work in the broader context of the literature, we point out three main findings. Our first was that \emph{company reviews contain linguistic markers of four stress types}. Previous work found that stress of social media users can be detected by analyzing their textual content, both on Twitter and Reddit~\cite{guntuku2019understanding}. Another study by Saha and De Choudhury found that high levels of stress markers were present in the use of language in Reddit comments posted by university students who experienced gun violence events at their campuses. This work showed that such linguistic changes are sufficiently subtle to reflect four \emph{different types of stress}, that is, low, passive, positive, and negative stress. Our second finding was that \emph{stress over the years tallied with large-scale exogenous events}. In particular, negative stress was the most prevalent among the four stress types in recession years (both great and mini recessions). This finding is in line with the literature linking economic downturn with stress and mental health issues caused by job instability ~\cite{mehta2015depression}, and speaks to the presence of linguistic markers reflecting negative stress associated with country-level economic performance. Our third finding was that \emph{company stock growth is associated with positive stress.} This is a new finding, not least because of lack of data in the past. While stock growth conflates endogenous factors (e.g., productivity) with exogenous ones (e.g., financial cycles), we found that positive stress companies enjoyed significantly stronger stock growth.

However, more work is needed to understand how to change a company's culture into one in which stressors could be used for one's growth and self-development. Given the recent wave of Great Resignation (i.e., the elevated rate at which U.S. workers have quit their jobs~\cite{economist21}), questions relating to corporate culture~\cite{economist22} and ways of retaining top talent are of utmost importance. A recent study from Mercer, an American asset management firm, found that elevated levels of employee turnover are not due to lack of engagement at work but attributed to workplace culture and heightened stressors. Therefore, organizations need to take immediate actions by (re)assessing their workplace culture first and by then shifting it when deemed appropriate, through training that fosters psychological safety and cultivates one's mindset towards positive stress. Traditionally, employee well-being has been tracked with tailored surveys. Automated analyses of the language used by employees on corporate social-networking tools might offer yet another way of tracking workplace stress, which is sufficiently granular to assess the impact of interventions in a company. Beyond the immediate use of these findings for individual companies, several other stakeholders could benefit from our methodology including government officials. As the performance of the S\&P 500 companies affects the broader U.S. economy, recommended workplace practices could be established at state- or national-level to improve work conditions.

\section*{Data availability}

We made our code and data available in a readily usable format on GitHub (\url{https://github.com/sanja7s/positive_stress_in_companies}) to allow for reproducibility. For each company, we shared the following attributes: \texttt{company name}, \texttt{\#total reviews}, \texttt{\#stress reviews}, \texttt{average rating}, \texttt{rating of work-life balance}, \texttt{rating of career prospects}, \texttt{rating of the company}, \texttt{rating of the culture}, \texttt{rating of the management}, \texttt{stress type}, \texttt{strength of association with the stress type}, \texttt{stock values/growth for: 2009, 2012, 2014, 2019}, and \texttt{industry sector}.

\bibliography{main}




\newpage

\section*{Supplementary Information}

\begin{figure*}[h!]
    \centering
    \includegraphics[width=0.65\linewidth]{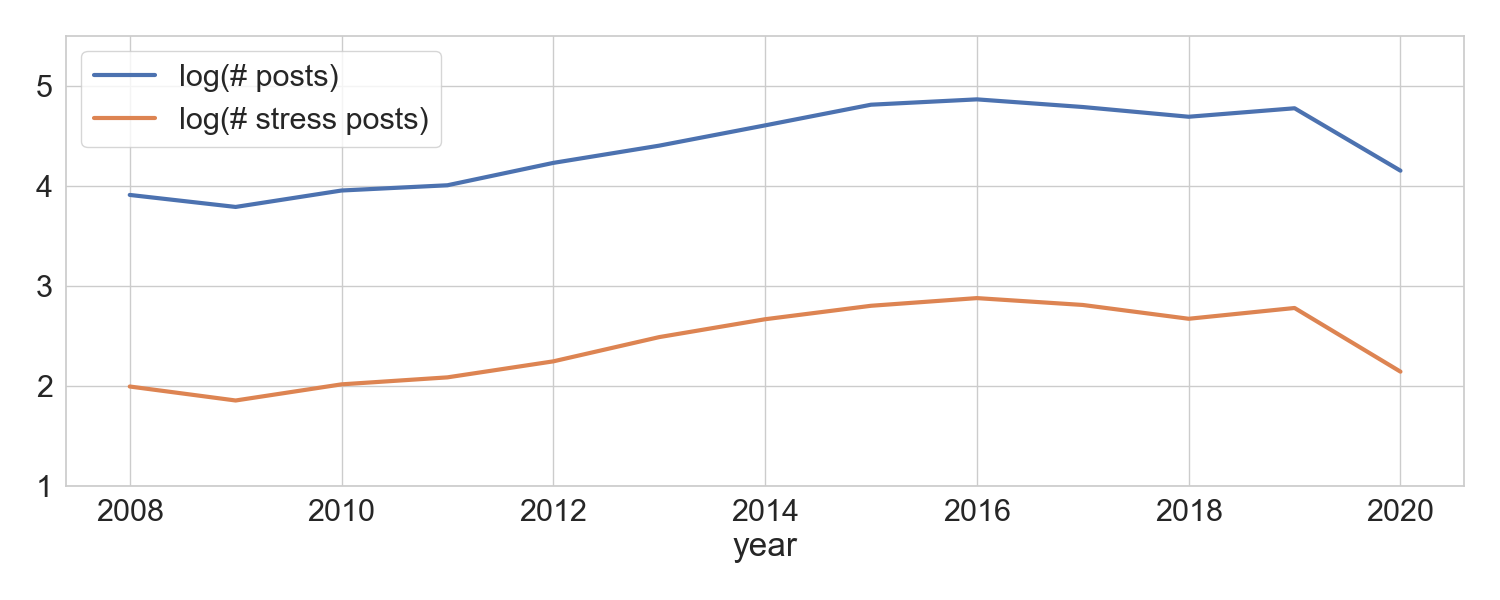}
    \caption{Number of posts (log) in our dataset between 2008 and 2020s. There is no trend difference between all posts and those containing mentions related to stress.}
    \label{fig:supp-temporal_reviews}
\end{figure*}

\section*{Details of the dataset}
We collected a total of 713,018 posts published for a company reviewing site from the start of 2008 up until the first quarter of 2020 for the S\&P 500 companies. We filtered out posts belonging to non-US based companies, yielding a total of 439,163 posts across $399$ unique S\&P 500 companies. The average rating across companies ranged from a minimum value of $1.62$ up to a maximum value of $5$ ($ \mu = 3.37, \sigma = 0.40)$. While the overall fraction of stress posts per company was $1.11\%$, this value ranged from $0\%$ up to $9.52\%$ across companies ($ \mu = 1\%, \sigma = 1\%)$.

\section*{Data representatives}

A total of 439,163 posts were analyzed. These posts are about companies distributed across all  the 51 U.S. states (Table~\ref{tabsup:posts_states}). The highest number of  posts were found in California (i.e., a total of 69,968 posts), while the lowest in Wyoming (i.e., a total of 222 posts). The posts span across 11 industries classified according to the Global Industry Classification Standard (GICS), with the highest number of posts for companies in Information Technology, and the least number in Real Estate (Table~\ref{tabsup:industries}). The posts were written by managers, sales associates, software engineers, analysts, among others (Table~\ref{tabsup:employees_titles}). Current employees make up 56\% of the reviews, and the remaining reviews are predominantly by former employees who held the job within the last five years.
The maximum annual number of posts between 2008 and 2020 was observed in 2016, while the lowest number of posts in 2009 (Figure~\ref{fig:supp-temporal_reviews}).

\begin{table}[h!]
\caption{Number of posts and number of offices on the company reviewing site across U.S. States, ranked by the number of posts published between 2008 and 2020 in descending order. The state of California had the most published posts, while the state of Wyoming had the least published posts. The Pearson correlation between the log number of posts and the number of companies per state in our data is $.98$, while the correlation between the log number of posts in our data and the log of population size across states is $.93$.}
\label{tabsup:posts_states}
\begin{minipage}{0.5\textwidth}
\centering
\begin{tabular}{ccc}
\toprule
\textbf{U.S. State} & \textbf{\# posts} & \textbf{\# offices} \\
\midrule
   CA &              69968 &      340 \\
   TX &              43629 &      342 \\
   NY &              37515 &      313 \\
   IL &              25157 &      290 \\
   FL &              24082 &      283 \\
   GA &              17888 &      275 \\
   WA &              15672 &      239 \\
   NC &              14072 &      268 \\
   PA &              14064 &      271 \\
   OH &              12447 &      263 \\
   MA &              12355 &      253 \\
   AZ &              11834 &      228 \\
   NJ &              11561 &      245 \\
   VA &              11320 &      235 \\
   CO &               9408 &      249 \\
   MN &               8437 &      196 \\
   MI &               7953 &      237 \\
   MO &               7657 &      230 \\
   TN &               7165 &      221 \\
   OR &               6704 &      206 \\
   MD &               6610 &      207 \\
   IN &               5727 &      222 \\
   WI &               5006 &      181 \\
   CT &               4567 &      186 \\
   KY &               4224 &      190 \\
   UT &               3925 &      187 \\
\bottomrule
\end{tabular}

\end{minipage} \hfill
\begin{minipage}{0.5\textwidth}
\begin{tabular}{ccc}
\toprule
\textbf{U.S. State} & \textbf{\# posts} & \textbf{\# offices} \\
\midrule
   OK &               3772 &      174 \\
   DC &               3596 &      180 \\
   KS &               3459 &      169 \\
   SC &               3362 &      194 \\
   NV &               2815 &      163 \\
   LA &               2631 &      175 \\
   AL &               2546 &      171 \\
   DE &               2067 &      113 \\
   IA &               1849 &      133 \\
   RI &               1676 &      103 \\
   AR &               1666 &      149 \\
   NH &               1627 &      127 \\
   NE &               1564 &      139 \\
   ID &               1208 &      112 \\
   MS &               1138 &      127 \\
   NM &               1097 &      125 \\
   WV &                803 &      114 \\
   ME &                604 &      105 \\
   HI &                600 &       85 \\
   ND &                344 &       74 \\
   VT &                324 &       54 \\
   MT &                322 &       70 \\
   AK &                300 &       57 \\
   SD &                297 &       57 \\
   WY &                222 &       72 \\
    &                 &        \\
\bottomrule
\end{tabular}
\end{minipage}
\end{table}

\begin{figure*}
    \centering
    \includegraphics[width=0.55\linewidth]{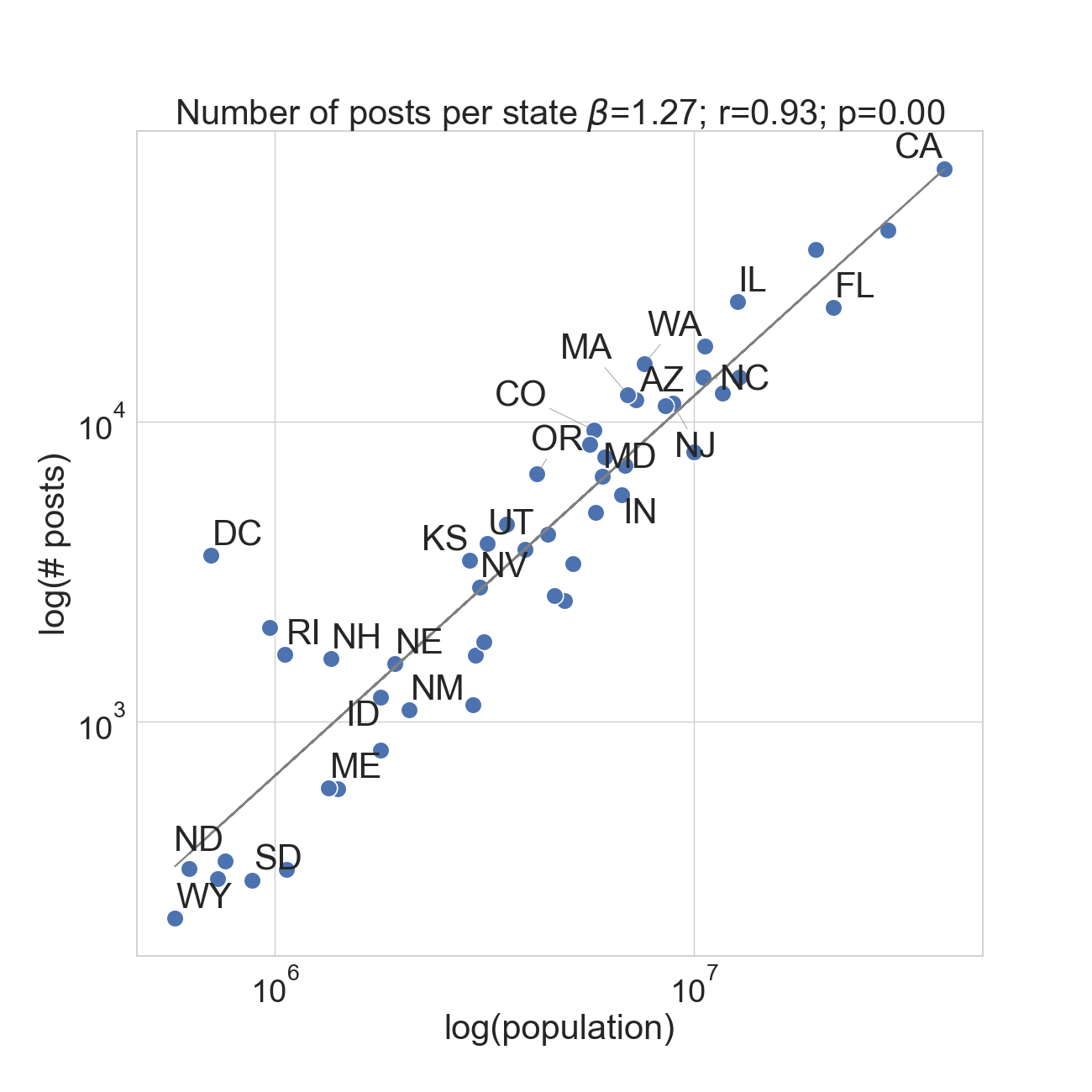}
    \caption{Number of posts (log) in our dataset versus state population (log). The  states Washington DC and Rhode Island have more posts that what the population size would suggest. The line of best linear fit is shown in gray. U.S. states are shown with the two-code state abbreviation.}
    \label{fig:supp-spatial_rep}
\end{figure*}

\begin{table}[t!]
\caption{Number of posts across the Global Industry Classification Standard (GICS) sectors. More posts are generally found in sectors having more companies, as one expects.}
\label{tabsup:industries}
\centering
\begin{tabular}{lrr}
\toprule \textbf{GICS Sector} &  \textbf{\# posts} &  \textbf{\# companies} \\
\midrule
 Information Technology &              63198 &         52 \\
 Consumer Discretionary &              62395 &         40 \\
             Financials &              49955 &         42 \\
            Health Care &              36308 &         41 \\
       Consumer Staples &              26471 &         28 \\
            Industrials &              24074 &         43 \\
 Communication Services &              13842 &         13 \\
                 Energy &               5510 &         20 \\
              Materials &               5269 &         21 \\
              Utilities &               3172 &         19 \\
            Real Estate &               2228 &         16 \\
\bottomrule
\end{tabular}
\end{table}

\begin{table}[t!]
\caption{Number of posts across roles and statuses.}
\label{tabsup:employees_titles}
\centering
\begin{tabular}{lr}
\toprule \textbf{Employee Title} &  \textbf{\# posts} \\
\midrule
                  Sales Associate &        8006 \\
                          Manager &        4536 \\
                Software Engineer &        4191 \\
  Customer Service Representative &        4058 \\
                          Cashier &        3726 \\
                         Director &        2819 \\
                  Project Manager &        2365 \\
                   Senior Manager &        2225 \\
         Senior Software Engineer &        2019 \\
                        Associate &        1969 \\
                    Store Manager &        1963 \\
                Assistant Manager &        1930 \\
              Pharmacy Technician &        1747 \\
                          Analyst &        1660 \\
                  Delivery Driver &        1619 \\
\toprule \textbf{Employee Status} &  \textbf{\# posts} \\
\midrule

Current Employee  &      190876 \\
Former Employee  &      146449 \\
Former Intern  &        5207 \\
Former Contractor  &        3380 \\
Current Intern  &        2858 \\
\bottomrule
\end{tabular}
\end{table}

\begin{figure*}
    \centering
    \includegraphics[width=0.45\linewidth]{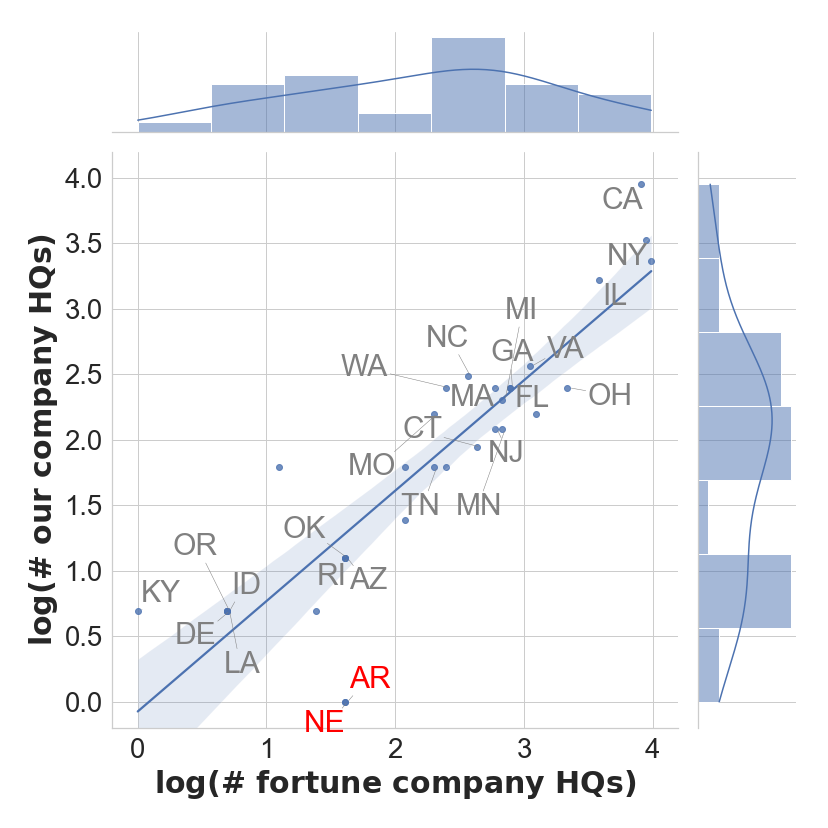}
    \caption{Correlation between the number of headquarters in each state in the Fortune 500 list and the number of headquarters in each state in our dataset (Spearman $r=.90$). The states of Nebraska (NE) and Arizona (AR) have fewer headquarters than what the Fortune list would suggests. 
    }
    \label{fig:supp-headquarters}
\end{figure*}

\newpage

\section*{Description and evaluation of the deep-learning framework}
To extract stress mentions, we used the MedDL entity extraction module~\cite{scepanovic2020extracting} (the left rectangle in Figure~\ref{fig:quadrant}(a)). MedDL uses \emph{contextual embeddings} and a \emph{BiLSTM-CRF sequence labeling architecture}. The BiLSTM-CRF architecture~\cite{huang2015bidirectional} is the deep-learning method commonly employed for accurately extracting entities from text~\cite{strakova2019neural,akbik2019pooled}, and consists of two layers. The first layer is a BiLSMT network (the dashed rectangle in Figure~\ref{fig:quadrant}(a)), which stands for Bi-directional Long Short-Term Memory (LSTM). The outputs of the BiLSTM are then passed to the second layer: the CRF layer (enclosed in the other dashed rectangle). The predictions of the second layer (the white squares in Figure~\ref{fig:quadrant}(a)) represent the output of the entity extraction module. To extract the medical entities of symptoms and drug names, BiLSTM-CRF takes as input representations of words (i.e., embeddings). The most commonly used embeddings are Global Vectors for Word Representation (GloVe) \cite{pennington2014glove} and Distributed Representations of Words (word2vec) \cite{mikolov2013distributed}. However, these do not take into account a word's context. The word `pressure', for example, could be a stress symptom at the workplace (e.g., \textit{`I felt constant \textbf{pressure} to deliver results'}) or could be used in the physics context (e.g., \textit{`The solid material found in the centre of some planets at extremely high temperature and \textbf{pressure}'}). To account for context, \emph{contextual embeddings} are generally used. MedDL used the RoBERTa embeddings as it had outperformed several others contextual embeddings, including ELMo, BioBert and Clinical BERT \cite{scepanovic2020extracting}.

\begin{figure*}
    \centering
    \includegraphics[width=0.35\linewidth]{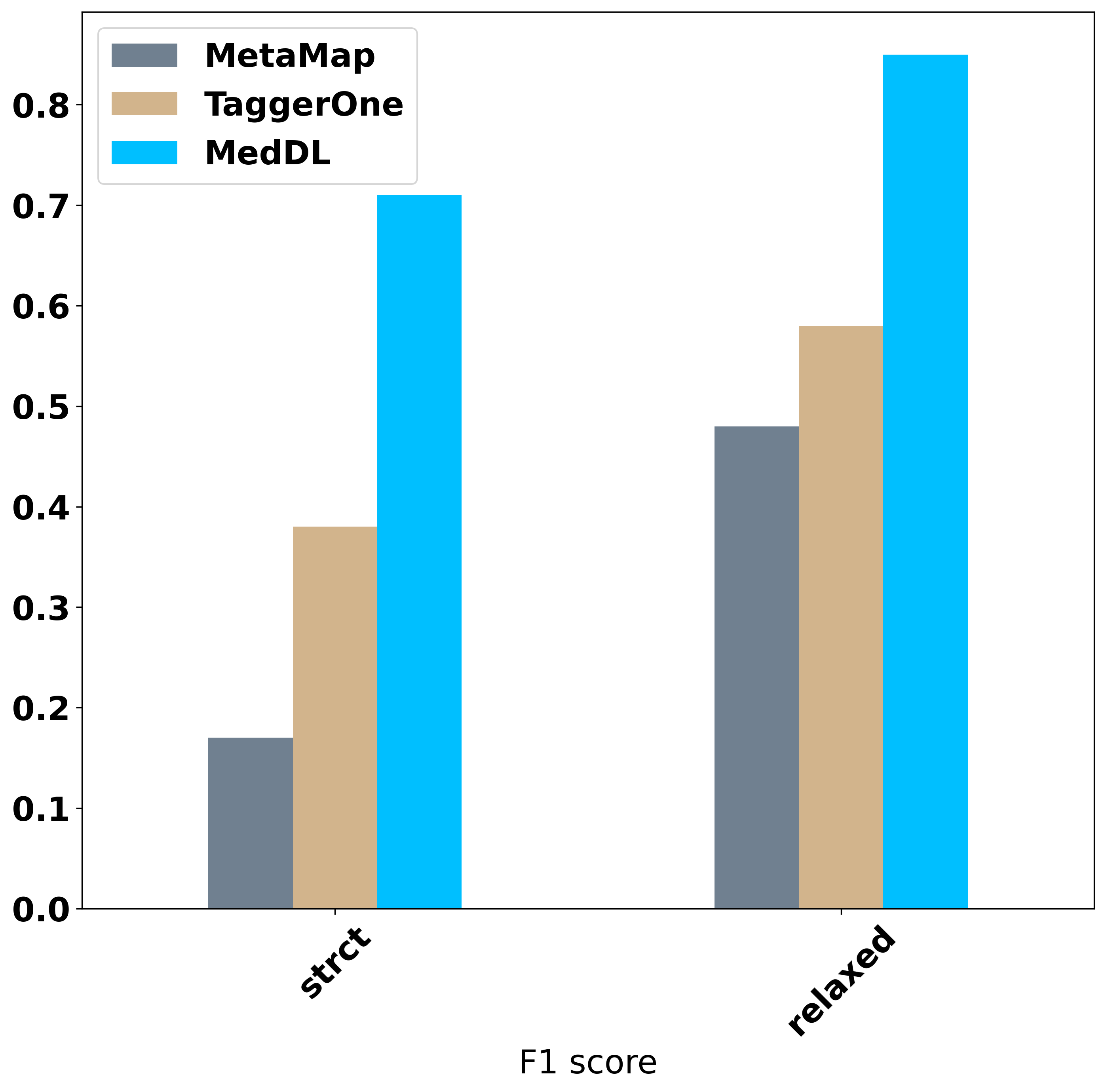}
    \caption{MedDL strict/relaxed F-1 score results when extracting medical symptoms on the MedRed dataset compared to two competitive alternatives of MetaMap and TaggerOne. }
    \label{fig:supp-medred}
\end{figure*}

Our evaluation metric is $F1$ score, which is the harmonic mean of precision $P$ and recall $R$:  
\begin{equation}
F1 = 2 \frac{P \cdot R} { P + R }
\end{equation}
\begin{equation*}
P=\frac{\# \textrm{correctly classified medical entities}}{\# \textrm{total  entities classified as being medical}}
\end{equation*}
and
\begin{equation*}
R=\frac{\# \textrm{correctly classified medical entities}}{\# \textrm{total medical entities}}.
\end{equation*}
For strict F-1 score, we counted as ``correctly classified'' only the entities that were \textit{exactly} matching the ground truth labels. For relaxed version of F-1 score, partially matching entities are also counted as correctly classified (e.g., if the model extracts the entity ``\emph{pain}'' given the full mention of ``\emph{strong pain}''). Also, given that our data comes with class imbalance (i.e., text tokens do not correspond equally to symptoms, or non-medical entities), we corrected for that by computing $P$ and $R$ using micro-averages~\cite{ash13}. In so doing, we were able to compare Med-DL's F1 scores with those  of two well-known entity extraction tools: MetaMap and TaggerOne. MetaMap is a well-established tool for extracting medical concepts from text using symbolic NLP and computational-linguistic techniques \cite{aronson2010overview}, and has become a de-facto baseline method for NLP studies related to health \cite{tutubalina2018medical}. TaggerOne is a machine learning tool using semi-Markov models to jointly perform two tasks: entity extraction and entity normalization. The tool does so using a medical lexicon \cite{leaman2016taggerone}. The MedDL pre-trained model was evaluated on a labeled dataset of Reddit posts called MedRed. The MedRed dataset was split into train (50\%), dev (25\%), and test (25\%) sets. The MedDL method achieved a strict/relaxed $F1$-score of $.71$/$.85$ when extracting symptoms (Figure SI \ref{fig:supp-medred}), outperforming both MetaMap and TaggerOne by a large margin (the two have $F1$-scores of $.17/.48$ and $.31/.58$, respectively).

 \begin{figure}
	\begin{center}
		\centering
		\begin{minipage}[b]{0.3\linewidth}
		    \centering
		    \includegraphics[width=1.0\linewidth]{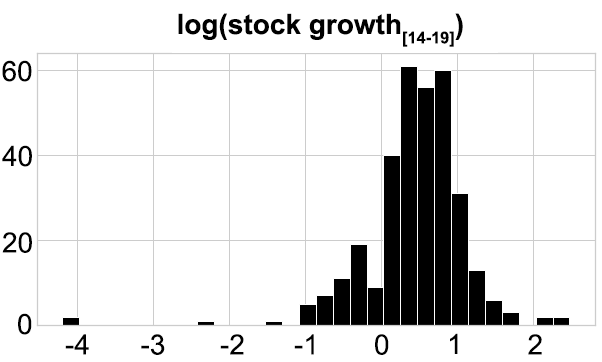}
	
		    \centerline{\small(a)}
		\end{minipage}
		\begin{minipage}[b]{0.6\linewidth}
		    \centering
		    \includegraphics[width=1.0\linewidth]{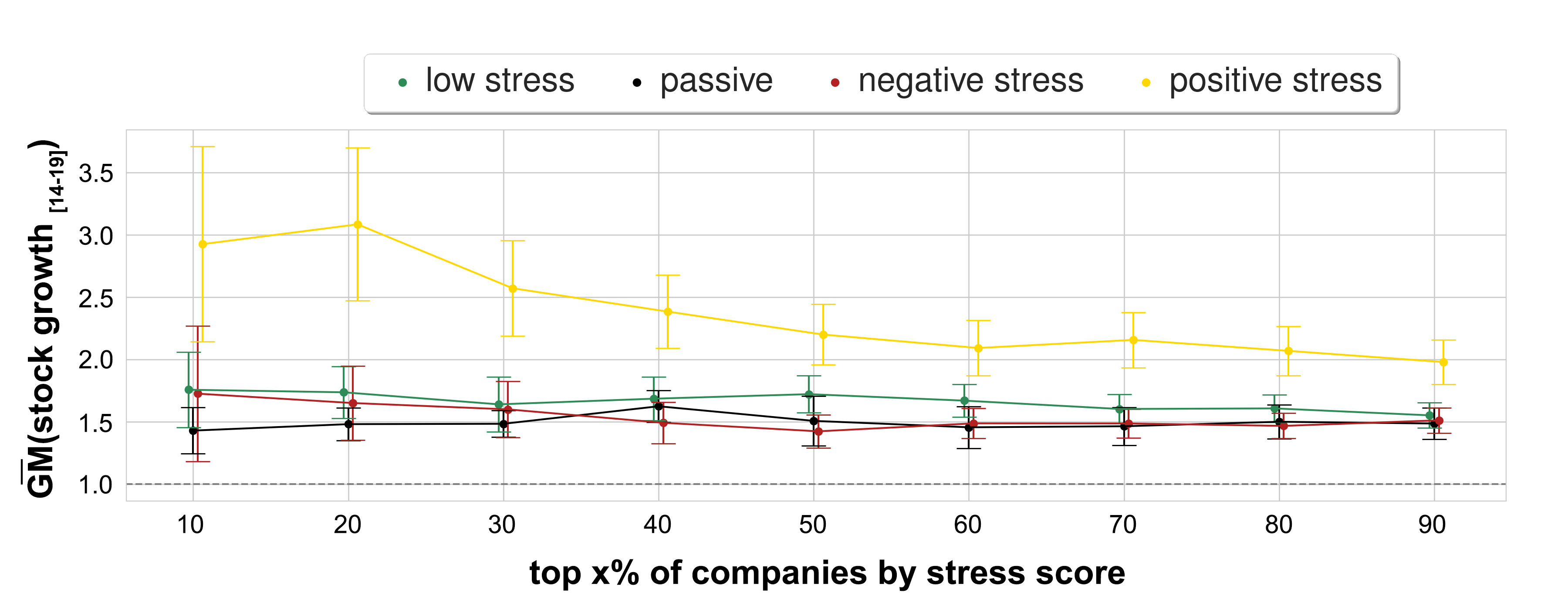}
		    \centerline{\small(b)}
		\end{minipage}
	\end{center}    
	\caption{ \emph{(a)} Distribution across companies of the logarithm of stock growth values from the average stock price in 2014 and that of 2019 (${stock\_growth}_{[14-19]} = stock_{2019}/stock_{2014}$) showing the stock growth is log-normally distributed. The average stock price for year $y$ ($stock_y$) is calculated as the average of the daily Adjusted Closing Prices for the year. \emph{(b)} Geometric mean of the stock growth values $\bar{GM}({stock\_growth}_{[14-19]})$ for increasing stress score percentiles for the companies of a given stress type. Error bars represent geometric standard error $GSE({stock\_growth}_{[14-19]}) =$ $\bar{GM}({stock\_growth}_{[14-19]})/$ $\sqrt{N} \cdot \sigma(log({stock\_growth}_{[14-19]}))$.}
	\label{fig:supp-stock_diff_14_19}
\end{figure}

\begin{figure*}[t!]
     \centering
    \includegraphics[width=.65\linewidth]{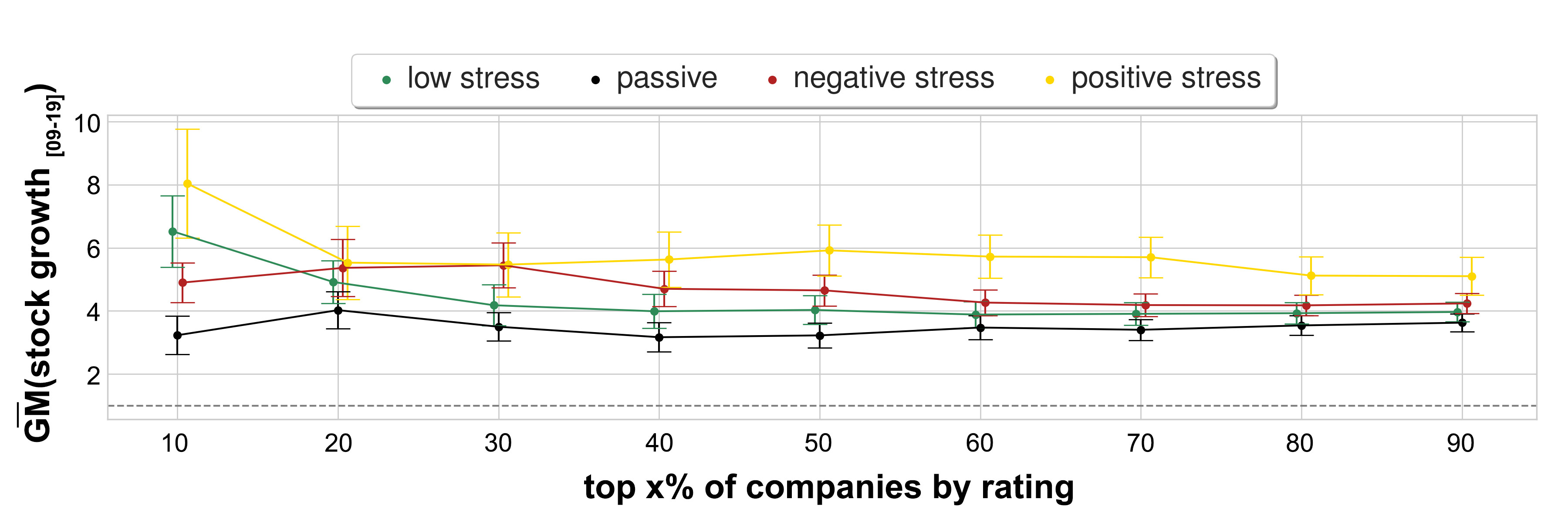}
     \caption{Geometric mean of the stock growth values $\bar{GM}({stock}\mbox{ }{growth}_{[09,19]})$ for different ratings percentiles for companies of the four stress types. Error bars represent geometric standard error $GSE({stock}\mbox{ }{growth}_{[09,19]}) =$ $\bar{GM}({stock}\mbox{ }{growth}_{[09,19]})/\sqrt{N} \cdot \sigma(log({stock}\mbox{ }{growth}_{[09,19]}))$.}
      \label{fig:supp-stock_growth_pct_rating} 
\end{figure*}

\newpage

 \begin{figure*}
	\begin{center}
		\centering
		\begin{minipage}[b]{0.49\linewidth}
		    \centering
		    \includegraphics[width=1.0\linewidth]{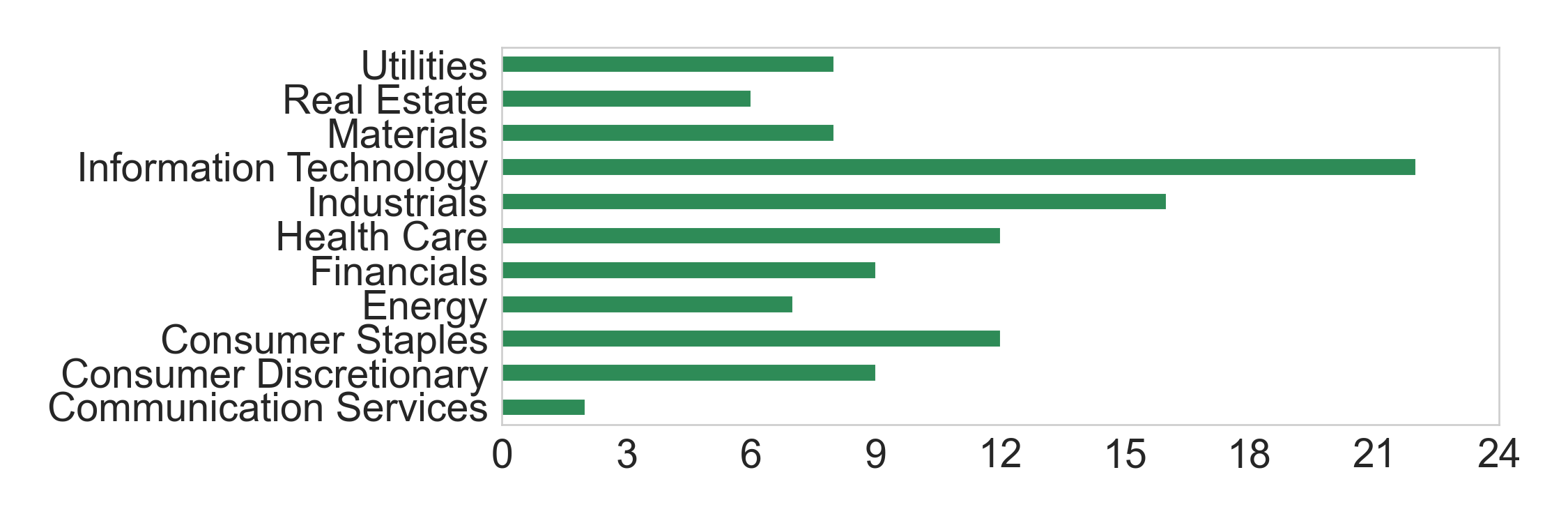}
		    \centerline{\small(a) Low stress}
		\end{minipage}
		\begin{minipage}[b]{0.49\linewidth}
		    \centering
		    \includegraphics[width=1.0\linewidth]{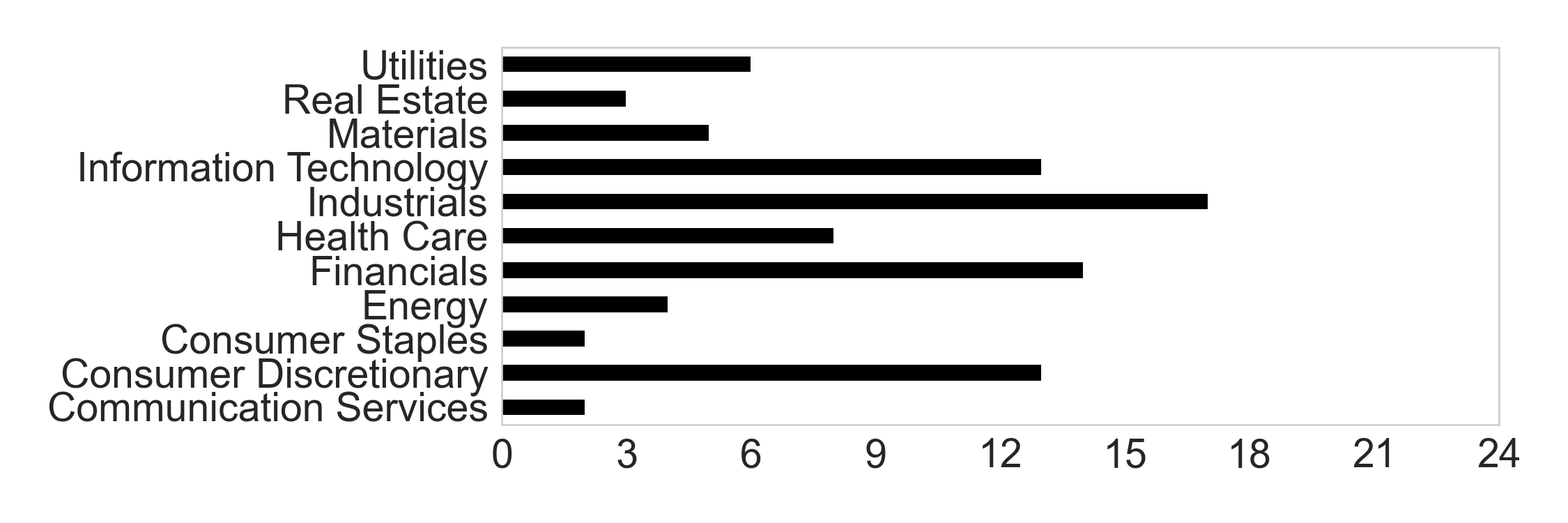}
		    \centerline{\small(b) Passive}
		\end{minipage}
		\newline
		\begin{minipage}[b]{0.49\linewidth}
		    \centering
		    \includegraphics[width=1.0\linewidth]{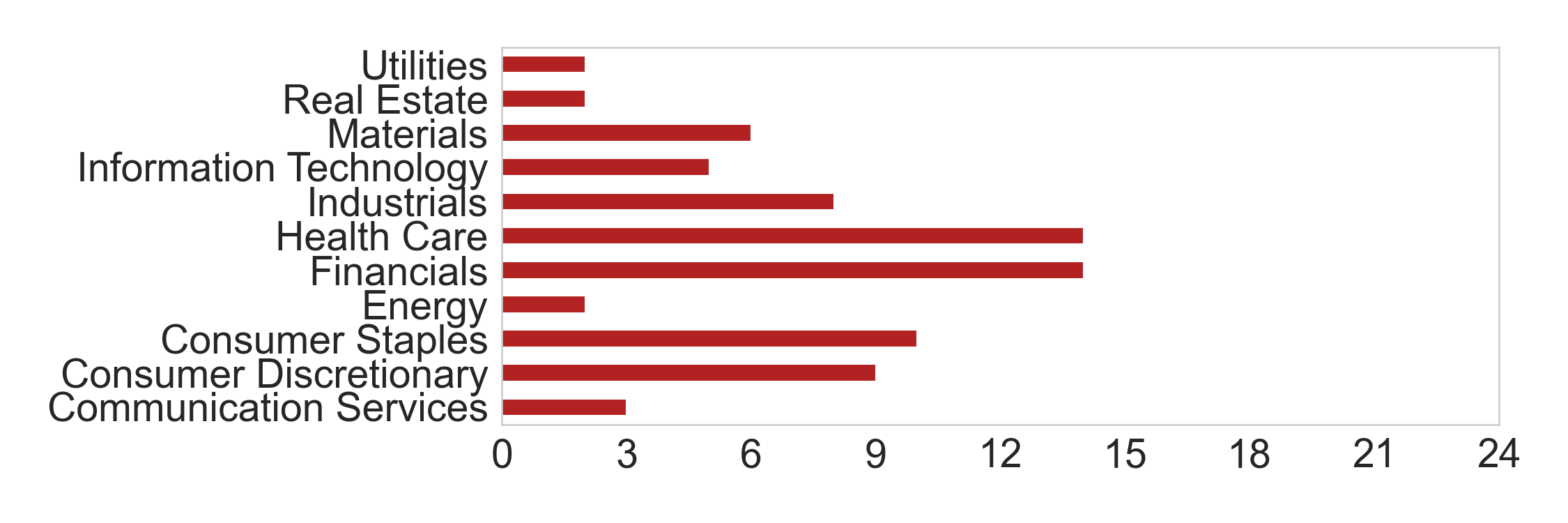}
		    \centerline{\small(c) Negative stress}
		\end{minipage}
		\begin{minipage}[b]{0.49\linewidth}
		    \centering
		    \includegraphics[width=1.0\linewidth]{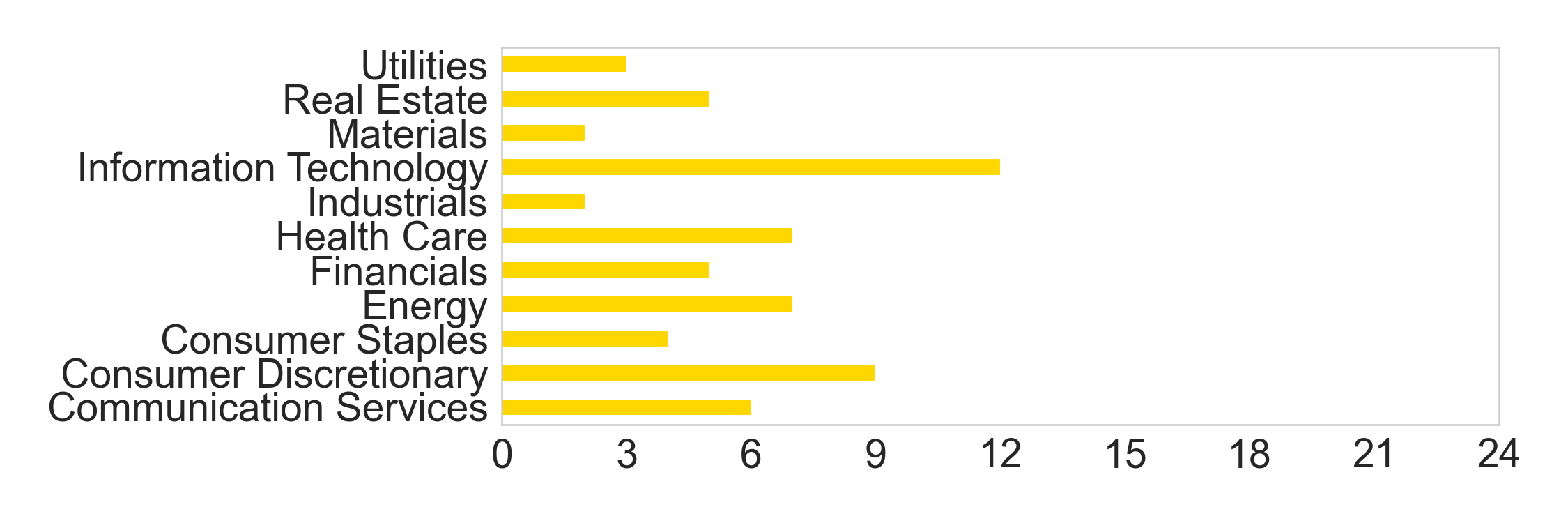}
		    \centerline{\small(d) Positive stress}
		\end{minipage}
	\end{center}    
	\caption{The number of companies per industry sector for the four stress types.  IT is more prominent among positive stress companies, while Health Care among negative stress companies.  }
    \label{fig:dist_industry_sectors}
\end{figure*}

\begin{figure}
	\begin{center}
		\centering
		\begin{minipage}[b]{0.7\linewidth}
		    \centering
		    \includegraphics[width=1.0\linewidth]{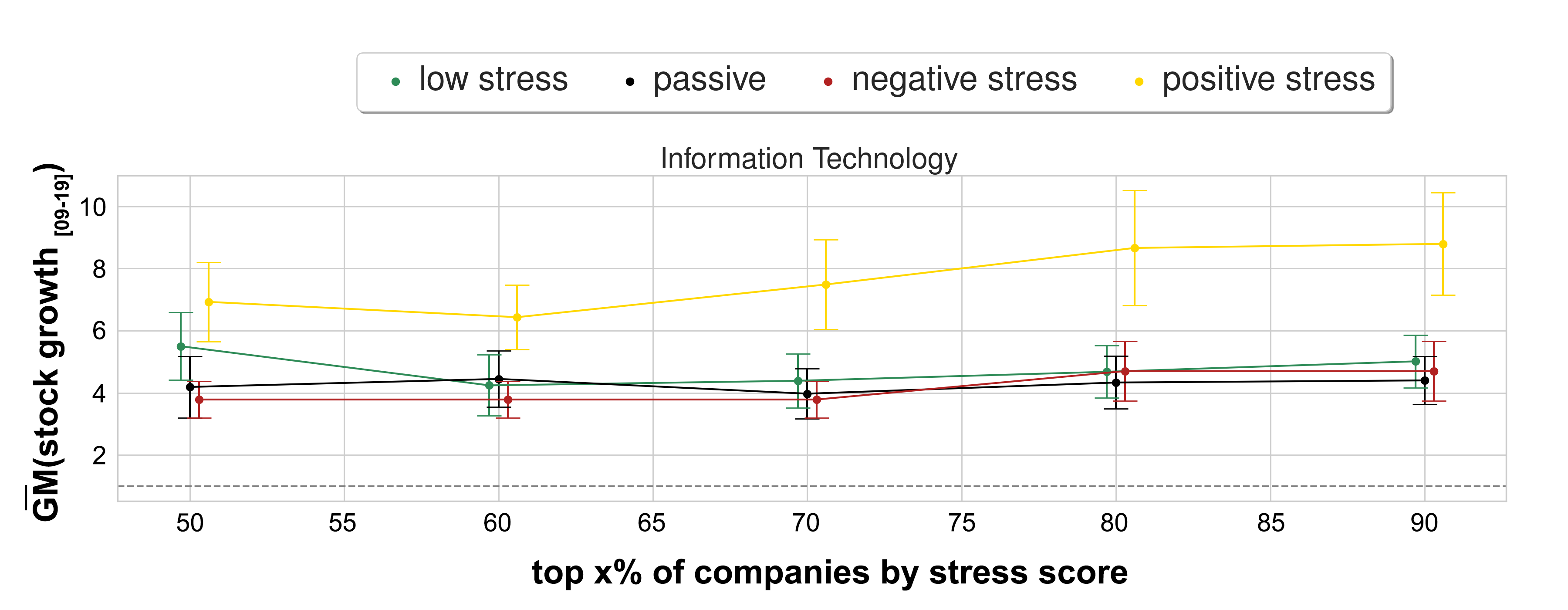}
		    \centerline{\small(a)}
		\end{minipage}
		\begin{minipage}[b]{0.7\linewidth}
		    \centering
		    \includegraphics[width=1.0\linewidth]{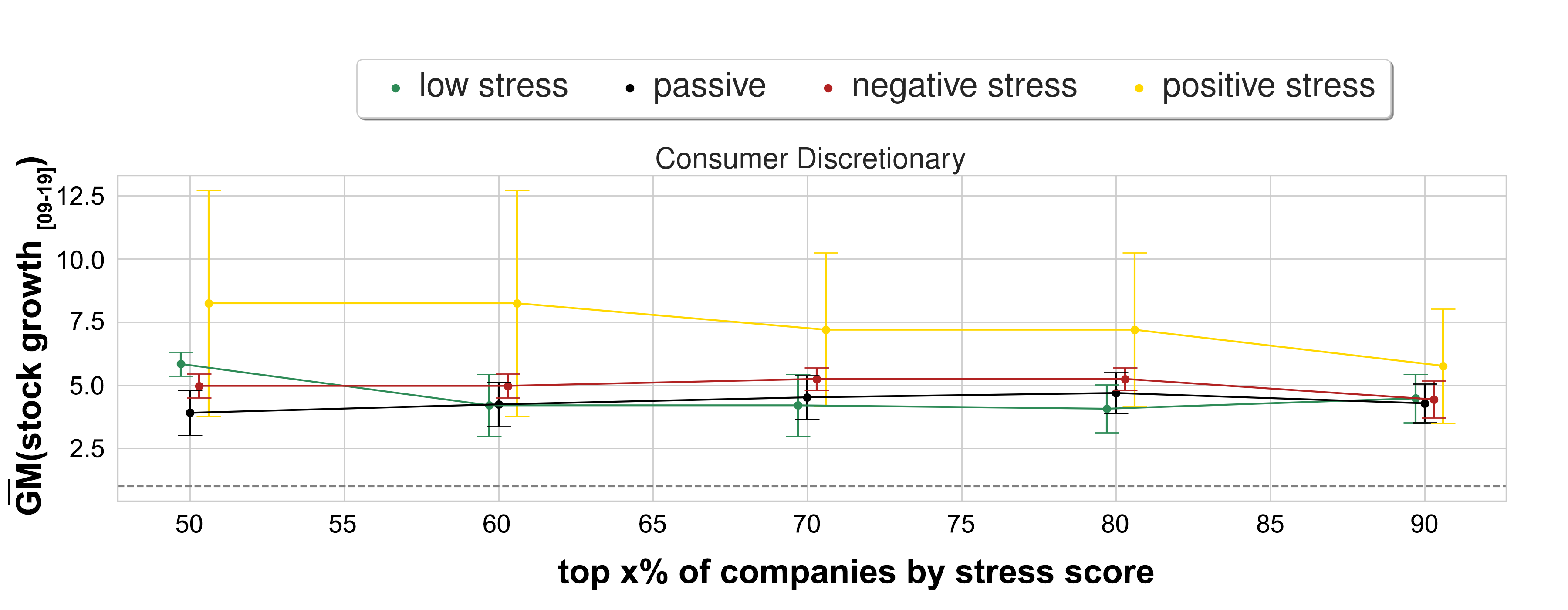}
		    \centerline{\small(b)}
		\end{minipage}
	\begin{minipage}[b]{0.7\linewidth}
		    \centering
		    \includegraphics[width=1.0\linewidth]{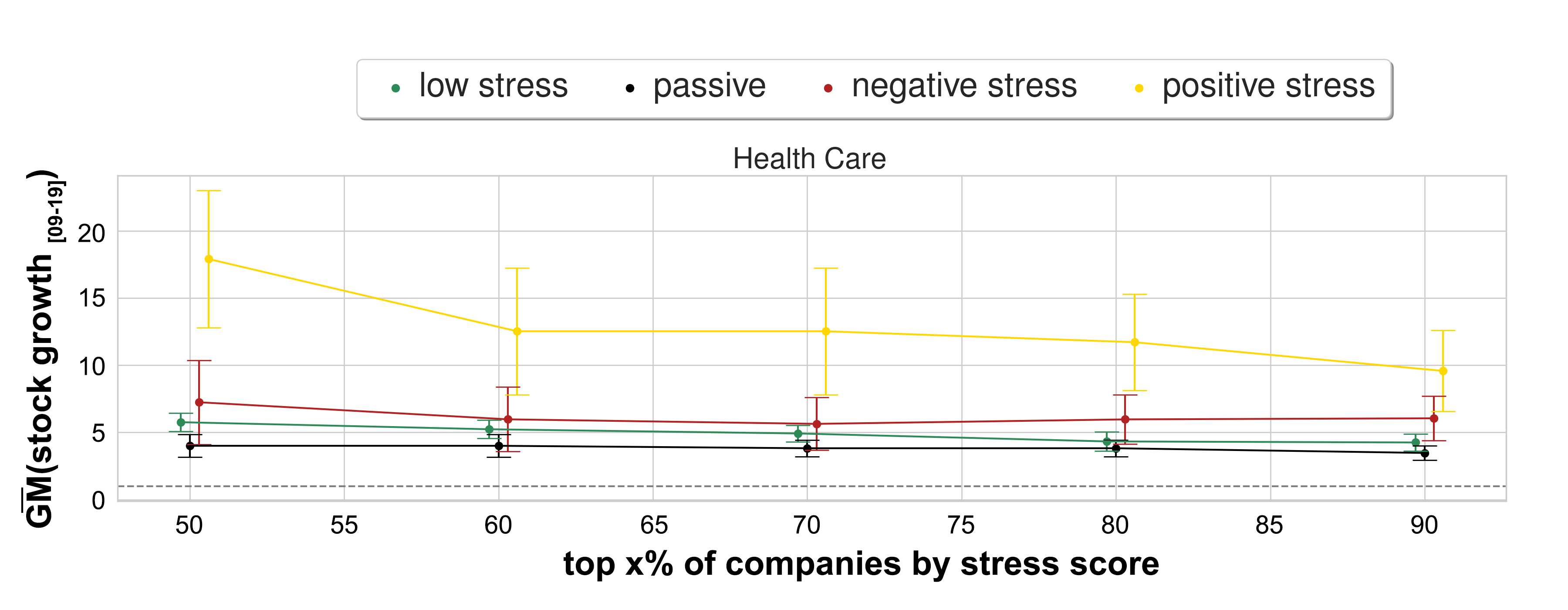}
		    \centerline{\small(c)}
		\end{minipage}
	\end{center}    
	\caption{Geometric mean of the stock growth values $\bar{GM}({stock\_growth}_{[09-19]})$ for increasing stress score percentiles for the companies in each of the three most present industry sectors: (a) Information Technology, (b) Consumer Discretionary, and (c) Health Care. The three sectors have sufficient data to ensure statistical significance for each percentile bin. Error bars represent geometric standard error $GSE({stock\_growth}_{[09-19]}) =$ $\bar{GM}({stock\_growth}_{[09-19]})/$ $\sqrt{N} \cdot \sigma(log({stock\_growth}_{[09-19]}))$.}
	\label{fig:industry_stock}
\end{figure}

\mbox{ } \\
\noindent
\textbf{Annotations of the words BERTopic found.} For each  topic, we identified the three most representative words and submitted the reviews mentioning them to six annotators. For example, we picked three reviews containing the words `overtime', `mandatory', and `shift' for negative stress companies, and asked six annotators to read them and describe what type of workplaces these reviews would suggest. Upon collecting a total of 72 free-form responses (i.e., each annotator described the reviews corresponding to the 12 topics), we conducted a thematic analysis~\cite{braun2006using}. To identify overarching themes, we used a combination of open coding and axial coding. We first applied open coding to identify key concepts. Specifically, one of the authors read the responses and marked them with keywords. We then used axial coding to identify relationships between the most frequent keywords to summarize them into semantically cohesive themes.

We found three high-level themes: \emph{career drivers}, \emph{industry or benefits}, and \emph{emotional aspects}. In the reviews, each theme was paraphrased differently depending on the four types of company stress, allowing us to identify sub-themes. The \emph{career drivers} theme described what motivated employees to go to work. Its sub-themes concerned companies whose employees experienced `considerable emotional pressure' (negative stress), tended to `focus on activities outside the work' (passive), cherished `their sense of control over their work' (low stress), and enjoyed `a collaborative and supportive workplace culture' (positive stress). In the \emph{industry or benefits} theme, we identified sub-themes mentioning either the industry sectors of the corresponding companies (e.g., Consumer Discretionary for negative stress, and Information Technology for positive stress) or  aspects concerning long-term financial benefits (e.g., passive and low stress). Finally, in the \emph{emotional aspects} theme, we identified sub-themes suggesting employees who experienced `emotional pressure' (negative stress), `tedious work' (passive), `good work-life balance' (low stress), or a `fast-paced, high-performing, and dynamic workplace environment' (positive stress).

\section*{Evaluation of BERTopic results}
We ran the topic modeling algorithm BERTopic~\cite{grootendorst10bertopic} separately on the four sets of reviews (each set containing reviews of the companies of a given stress type). The fact that BERTopic discovered distinct topics in the four sets reveals that stress is paraphrased differently in the sets. We calculated the topical overlapping values for the different combinations of the four sets (using the Jaccard similarity on the sets of keywords from the top ten topics of each stress type), and found them to be (on average) as low as 0.08 (on a scale ranging from 0 to 1).

\section*{Evaluation of the four quadrants}
\begin{figure*}[t!]
     \centering
    \includegraphics[width=.55\linewidth]{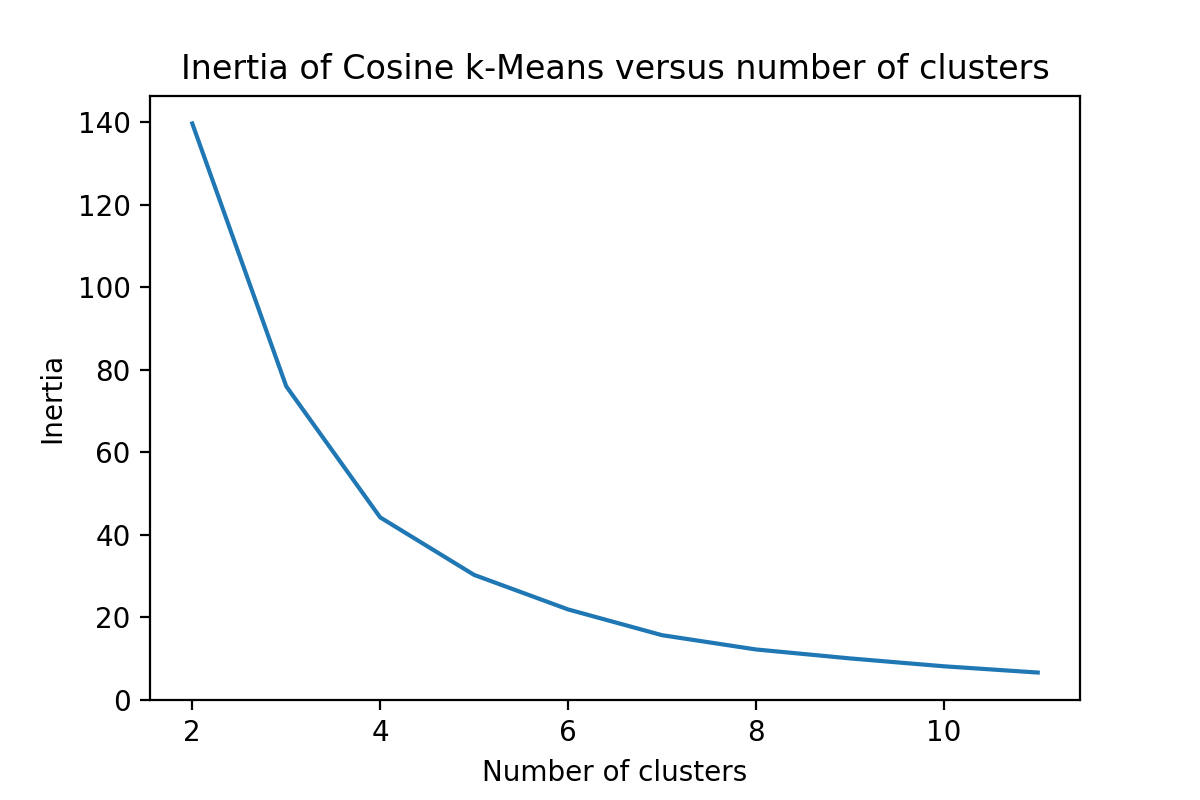}
     \caption{Inertia of Cosine k-Means versus number of clusters having the ``elbow'' at k=$4$.}
      \label{fig:supp:kmeans-elbow} 
\end{figure*}

To test whether the quadrant division of companies into four types was meaningful, we manually inspected 30 posts taken at random from companies with high stress, and found stress mentions in companies with low ratings to be qualitatively different from those in companies with high ratings (e.g., a review from a lowly rated company \emph{``The pressure is constantly high, while your work is not appreciated [...] and it feels like the managers do not know what they are doing.''} versus a review from a highly rated company \emph{``Happy Employee. Best culture I have experienced, especially in a stressful job. [...] The job is hard, but nothing worth having comes easy.''}). Similarly, we found qualitatively different review between companies with low stress and high versus low ratings (e.g., a review from a highly rated company \emph{``Solid company offering Work From Home. [...] decent options to choose for hours worked, great tech support, all equipment supplied, always feel connected to team, strong work ethic. ''},  versus a review from a lowly rated company \emph{``Sinking Ship due to Horribly Managed [...] Merger. At legacy X office, they managed to retain some of the positive company culture leftover from the X days. The people are still the best part of that office, but with the increasing turnover, layoffs and ``Hunger Games'' management style, that is in danger of ending... ''}).
    As a final validity check, we arranged companies along the two axes and clustered them in an unsupervised way. We found four to be the best number of clusters.
    More specifically, we applied k-means clustering, and searched for the optimal number of clusters using the elbow method (Figure \ref{fig:supp:kmeans-elbow}). The method involves calculating the sum of squared distances between data points and the $k$ assigned clusters' centroids, for an increasing number of clusters $k$. Once this value stops decreasing significantly, it means that that the optimal number of clusters is reached.
    
\section*{Sensitivity of the results}
\begin{figure*}[t!]
    \centering
        \includegraphics[width=.82\linewidth]{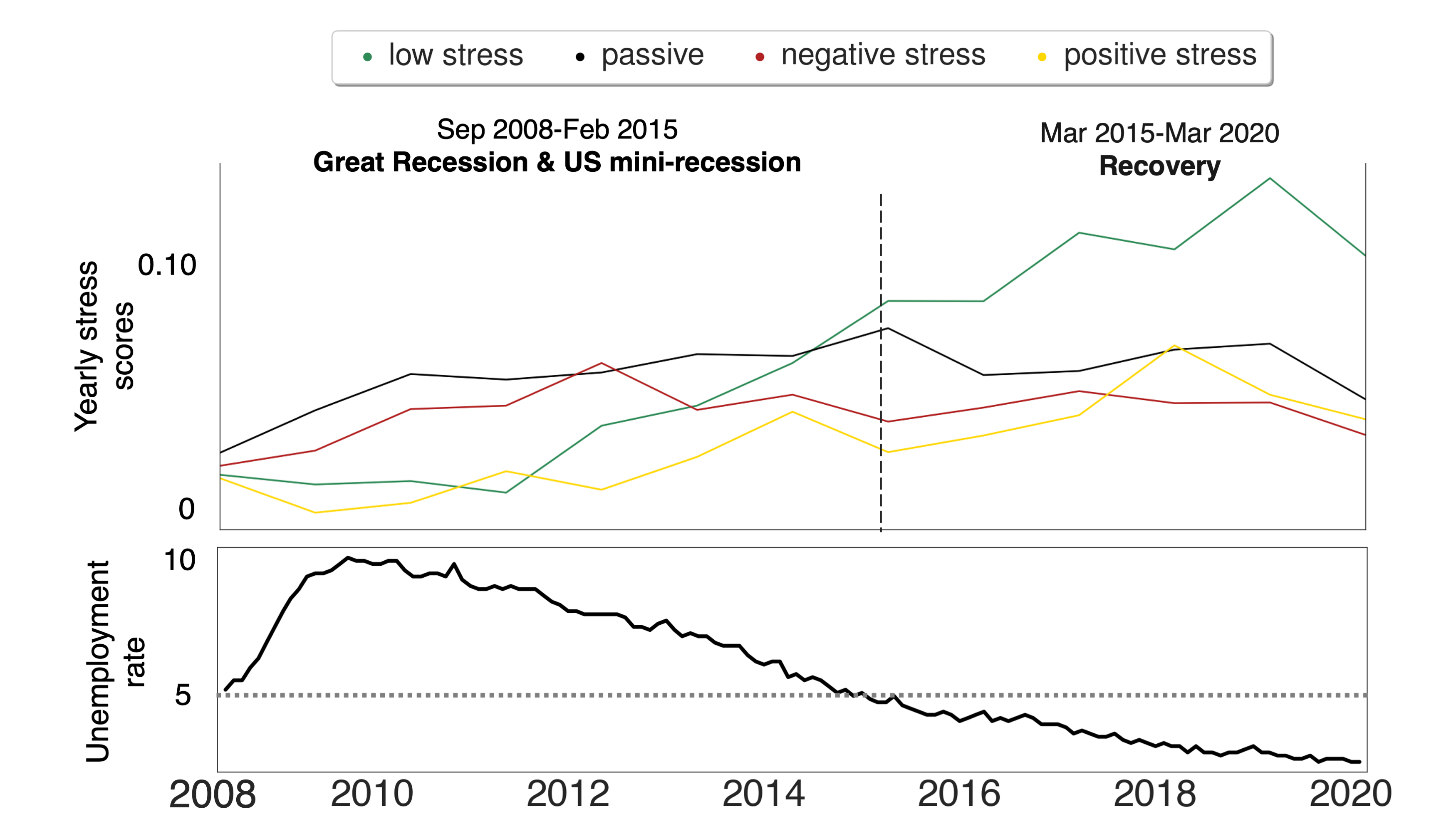} 
    \caption{The effects of weighting the yearly scores. \emph{(top)} The evolution of temporal scores without weights, i.e., where $w=1$ for the four types of stress; and \emph{(bottom)} the unemployment rate in the U.S., with  the horizontal dashed line reflecting pre-recession rate. The stress score per year  is calculated using Equation~(\ref{eq:q_temp_aggr}) with $w=1$.}
    \label{fig:supp:stress_time_eval}
 \end{figure*}

\mbox{ }\\
\textbf{Weighting the scores.} We explored the effects of weighting the yearly scores in:
\begin{align}\label{eq:q_temp_aggr}
    	m{(s,y)} = 
    	\sum_{c \in s} f(c,s,y) \times w{(c,y,s)},
\end{align}
by plotting the temporal scores without weights, i.e., where $w=1$. The result is shown in Figure \ref{fig:supp:stress_time_eval}. The simple aggregation skews the results towards (the long tail of) small companies as it considers a small company equal to a big one.

\mbox{ }\\
\textbf{Shorter-term growth.} To test whether our results on  stock growth are not affected by exogenous events such as the Great Recession,  we computed stock growth for the narrower 5-year period between 2014 to 2019:
\begin{equation}
{stock}\mbox{ }{growth}_{[14-19]}=  \frac{stock^{2019}}{stock^{2014}}
\end{equation}
where $stock_{i}$ is the average adjusted closing price of their stocks in year $i$. Figure~\ref{fig:supp-stock_diff_14_19} shows that the trend remains qualitatively the same as that in Figure 2, even when removing the Great Recession period. Positive stress companies enjoyed the highest stock growth (with average value across all percentiles being $\bar{GM}(\textrm{stock growth}_{[14-19]}) = 1.97$ as per Figure \ref{fig:supp-stock_diff_14_19} on the right), low stress companies had the second highest  ($\bar{GM}(\textrm{stock growth}_{[14-19]}) = 1.53)$, while passive and negative stress companies enjoyed the lowest growth ($\bar{GM}({\textrm{stock growth}_{[14-19]}}) = 1.46$, and  $1.45$, respectively). 

\mbox{ }\\
\textbf{Interaction effects between stress scores and review ratings.} We tested whether our observed stock growth was genuinely associated with positive stress companies rather than being simply associated with highly-rated companies. To this end, for each stress type, we plotted $\bar{GM}({stock\_growth}_{[09-19]})$ against different rating percentiles (Figure \ref{fig:supp-stock_growth_pct_rating}). Highly rated companies  experienced stock growth, yet there are still significant differences across companies of different stress types: in particular, positive stress companies of varying rating percentiles consistently enjoyed the highest growth (the yellow line in Figure~\ref{fig:supp-stock_growth_pct_rating} is consistently above the other three lines).

\mbox{ }\\
\textbf{Growth per industry sectors.} To test whether a specific industry sector is predominant for a given stress type, we first plotted the number of companies per industry sector according to the GICS classification  (Figure~\ref{fig:dist_industry_sectors}).
Information Technology was more prominent among positive stress and low stress companies, Health Care and Financials among negative stress ones, and Industrials and Consumer Discretionary among passive ones. To then check whether the distribution of industry sectors across the four types of stress affected our findings for stock growth, we computed stock growth between 2009 and 2019, and did so for the three most frequent industry sectors separately (i.e., Information Technology, Consumer Discretionary, and Health Care). We chose those three sectors because each individually contained a sufficient number of companies and, as such, allowed us to obtain statistical significant results.

Stock growth was computed as $GM({\textrm{stock growth}_{[09-19]}}) = \Pi (\textrm{stock growth}_{[09-19]}(c))^{1/n} $, where $c$ is each company from a given industry sector (e.g., Information Technology) in a specific \emph{(stress type,percentile)} bin, and $n$ is the number of the companies in such a bin.  For the three industry sectors, we plotted $\bar{GM}({stock\_growth})$ against different stress score percentiles (Figure \ref{fig:industry_stock}). In all three sectors, we observed that positive stress companies had consistently higher stock growth compared to the other three stress types.

\begin{figure}
     \centering

    \includegraphics[width=.8\linewidth]{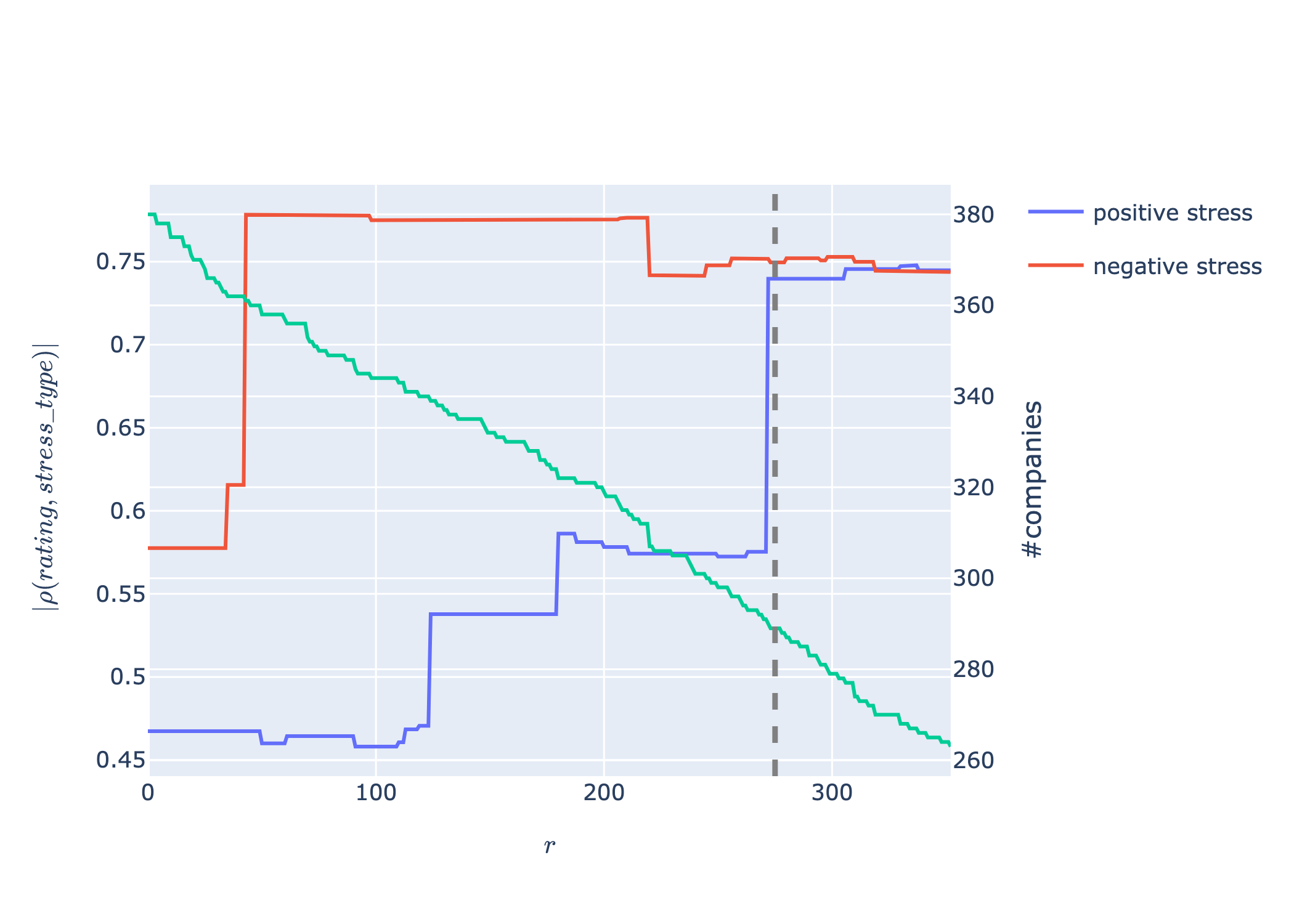}

    \caption{\textbf{Threshold selection.}  Correlation values between each of the two stress scores and a company's website overall rating ($y$-axis) for the companies with at least $r$ reviews ($x$-axis). These values have a phase shift at $r=280$ for positive stress companies (blue), matching the value of the correlation for negative stress companies (red).}
    \label{fig:th}
\end{figure}

\mbox{ }\\
\textbf{Percentage of stress posts.}
To test the sensitivity of our results to the percentage of stress posts being considered, we repeated our analyses by including only the companies with at least $r$ reviews. We found the optimal threshold $r$ to be $280$, and did so as follows.
To include at least half of the total S\&P 500 companies, the least number of reviews per company had to be less than $r=350$.
Then, for each $r = 1,...,350$, we subset the companies having at least $r$ reviews, and calculated the correlation between a company's rating and its positive stress score (for positive stress companies) or its negative stress score (for negative stress companies), and did so for each subset.
We found that the absolute values of the correlations increased with the number of reviews (Figure \ref{fig:th}), as one expected, and there was a phase shift at $r=280$ for positive stress companies ($\rho($company\_rating, positive\_stress\_association)=$.75$). The same applied to negative stress companies (Figure \ref{fig:th}).
At this threshold, we were left with $287$ companies out of $380$ companies in total. We repeated the calculations on this subset of companies and, compared to our previously reported results, found even stronger associations between: i) negative stress scores in the whole U.S. and the Great Recession, and ii) a company's positive stress score and its stock growth.

\begin{figure*}
  \centering
    \includegraphics[width=.8\linewidth]{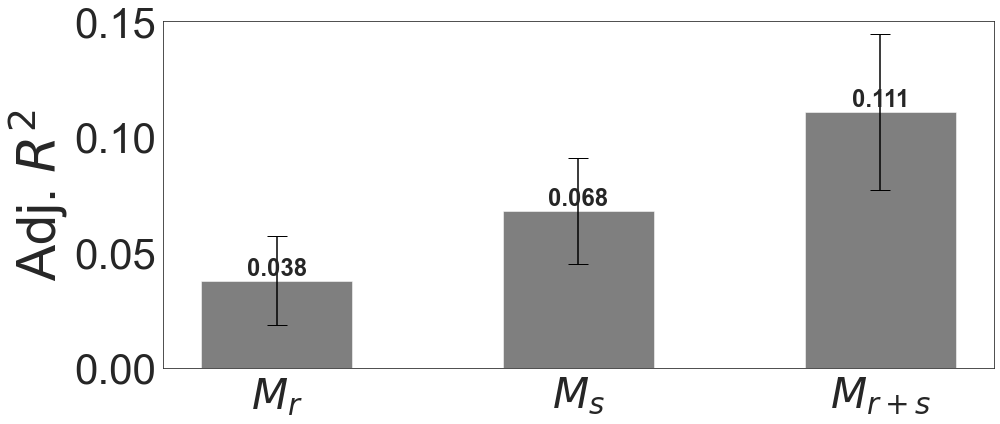}
    \caption{Adjusted $R^2$ values of three OLS models with different predictors: r is the rating score; s is the stress score; r+s is the rating score and stress score. We applied bootstrapping to ascertain the statistical significance of the results by randomly subsample a set of 120 companies 10 times. Average values and standard deviations are reported. We observed a 78\% and a 192\% increase in $M_{s}$ and in $M_{r+s}$ over the baseline model, respectively.}
    \label{fig:ols}
\end{figure*}

\mbox{ }\\
\textbf{Combining stress and review scores.} We fit three Ordinary Least Squares (OLS) models to predict stock growth (Figure \ref{fig:ols}). In each model, we used the (log of the) number of reviews as a control variable. In addition, (a) $M_{r}$ uses the average rating score as the additional independent variable (baseline model), (b) $M_{s}$ uses the stress score, and (c) $M_{r+s}$ uses both the rating score and the stress score as additional independent variables. We applied bootstrapping to ascertain the statistical significance of the results by randomly subsampling a set of 120 companies 10 times. We observed a 78\% and a 192\% increase in $M_{s}$ and in $M_{r+s}$ over the baseline model, respectively.

\end{document}